\documentclass[useAMS,usenatbib,fleqn]{mnras}
\usepackage{calrsfs}
\DeclareMathAlphabet{\pazocal}{OMS}{zplm}{m}{n}
\usepackage{multirow}
\usepackage{hhline}
\usepackage{times}
\usepackage{graphics,epsf}
\usepackage[T1]{fontenc}
\usepackage{aecompl} 
\usepackage{amsmath}               
\usepackage{placeins}
\usepackage{booktabs}
\usepackage{amsfonts}               % American Mathematical Society fonts
\usepackage{amssymb}                % American Mathematical Society symbol
\usepackage{epsfig}                 % EPS figures
\usepackage{epstopdf}
\usepackage{rotating}
\usepackage{color}
\usepackage{tabularx}
\usepackage{url}
\hypersetup{draft}

\def \apj{ApJ}
\def \aap{A\&A}
\def \aj{AJ}

\def \mnras{MNRAS}
\def \apjl{ApJ Lett.}
\def \apjs{ApJ Suppl. Ser.}
\def \nat{Nature}

\def \pasp{PASP}

\def \araa{ARA\&A}

\begin{document}

\title[Dynamics of mildly hierarchical triple systems]{Quasi-secular evolution of mildly hierarchical triple systems: \\
analytics and applications for GW-sources and hot Jupiters}

\author[Grishin et. al.]{
Evgeni Grishin,$^{1}$
Hagai B. Perets,$^{1}$
Giacomo Fragione $^{2}$
\\
$^{1}$Physics Department, Technion - Israel institute of Technology, Haifa,
Israel 3200002\\
$^{2}$Racah Institute for Physics, The Hebrew University, Jerusalem, Israel 91904\\
%\author[Grishin et. al.]{Evgeni  Grishin, Dong Lai \& Hagai B. Perets
%\\
E-mail: eugeneg@campus.technion.ac.il (EG); giacomo.fragione@huji.ac.il (GF); hperets@physics.technion.ac.il (HBP)}
%$^1$Physics department, Technion - Israel institute of Technology, Haifa, Israel 3200002
\maketitle

\begin{abstract}
In hierarchical triple systems, the inner binary is perturbed by a
distant companion. For large mutual inclinations, the Lidov-Kozai mechanism secularly excites large eccentricity and inclination oscillations of the inner binary. The maximal eccentricity attained, $e_{\rm max}$ is simply derived and widely used. However, for
mildly hierarchical systems (i.e. the companion is relatively close
and massive), non-secular perturbations affect the evolution. Here
we account for fast non-secular variations and
find new analytic formula for $e_{{\rm max}}$, in terms of the
system's hierarchy level, correcting previous work and reproducing
the orbital flip criteria. We find that $e_{{\rm max}}$ is generally
enhanced, allowing closer encounters between the inner binary components, thus significantly changing their interaction and its final outcome.
We then extend our approach to include additional relativistic
and tidal forces. Using our results, we show that the merger time
of gravitational-wave (GW) sources orbiting massive black-holes in
galactic nuclei is enhanced compared with previous analysis accounting
only for the secular regime. Consequently, this affects the distribution
and rates of such GW sources in the relevant mild-hierarchy regime.
We test and confirm our predictions with direct N-body and 2.5-level Post-Newtonian
codes. Finally, we calculate the
formation and disruption rates of hot-Jupiters (HJ) in planetary systems
using a statistical approach, which incorporates our novel results
for $e_{{\rm max}}$. We find that more HJ migrate from further out,
but they are also tidally disrupted more frequently. Remarkably, the
overall formation rate of HJs remains similar to that found in previous
studies. Nevertheless, the different rates could manifest in different underlying
distribution of observed warm-Jupiters.
\end{abstract}


\section{Introduction}

Three body systems are ubiquitous in astrophysics and appear in a
plethora of configurations and scales, from moons and asteroids of
planets, to multiple stars and binary compact object around supermassive
black holes. The general three body problem is notoriously non-integrable
\citep{1892mnmc.book.....P}, but some special cases allow useful
analytic approximations that shed light on their features \citep{ValtonenBook2006}.

Hierarchical triples are systems where an inner binary is perturbed
by a third distant companion. Observations of exo-planets \citep{2011PASP..123..412W,Kn14,2015ARA&A..53..409W},
multiple stars \citep{Rag10,2014AJ....147...86T} and compact objects
in extreme orbital inclinations and eccentricities call for better
understanding of such hierarchical multiple systems. The key parameter
in the study of evolution of hierarchical systems is the maximal eccentricity
$e_{{\rm max}}$ of the (inner) binary. Under appropriate conditions
large eccentricities can be induced in the inner binaries of hierarchical
triples through secular processes. These, in turn, can result in close
encounters of the inner binary components during their pericentre
approach, giving rise to a plethora of astrophysical phenonema, depending
of the astrophysical set-up. Such processes include tidal dissipation in triple stars
\citep{1998MNRAS.300..292K,2001ApJ...562.1012E}, Hot-Jupiter (HJ)
formation \citep{wu03,2007ApJ...669.1298F, Naoz11nat,Anderson_ps,PetHJ,munoz16},
secular evolution of planets and satellites \citep{Per+09,T9,Grishin17,Grishin18},
triple stellar evolution \citep{2009ApJ...697.1048P,perets12, hamers13, michaely14,frewen16, T16, stephan2018}, gravitational-wave
(GW) emission and mergers \citep{wen03, 2012ApJ...757...27A,ant14,AMK14,sil17,LiuLai17, LL18,Randall_LK2,Randall_LK1,FK18,hvrr},
tidal disruption events \citep{fraglei18a,fraglei18b}, direct
collisions and type Ia supernovae \citep{katz_sne} etc.

The main approach in studying the long-term evolution of hierarchical
triples is through a perturbative method. In hierarchical systems,
the interaction potential is expanded in multipoles in the (small)
ratio of the inner to outer separations, and than double-orbit-averaged
(DA) on both orbits \citep[and references therein]{Kozai62,Lidov62,Naoz2016review}
. The resulting leading DA quadrupole term is integrable and the system
admits an exact analytic solution \citep{2007CeMDA..98...67K}. Lidov-Kozai
(LK) oscillations occur if the mutual inclination is in the range
of the well known critical values $i_{c}=\arccos\pm\sqrt{3/5}=39.2^{\circ}, 140.8^{\circ}$.
During the LK cycle, the maximal eccentricity attained is 
\begin{equation}
e_{{\rm max}}^{{\rm DA}}=\sqrt{1-\frac{5}{3}\cos^{2}i_{0}}, \label{eq:emax}
\end{equation}
where $i_{0}$ is the initial mutual inclination, if the initial eccentricity
$e_{0}\ll1$ is low. Eq. (\ref{eq:emax}) can be derived using the
conservation of the specific $\hat{z}$ component of the inner binary's
angular momentum, $j_{z}=\sqrt{1-e^{2}}\cos i$ (in the limit where
the outer angular momentum dominates, i.e. the test particle limit),
where $e$ is the inner binary eccentricity. The typical (secular)
timescale for change in the orbital elements is \citep{2007CeMDA..98...67K, antognini15}
\begin{equation}
\tau_{{\rm sec}}\approx\frac{1}{2\pi}\frac{m_{{\rm tot}}}{m_{{\rm out}}}\frac{P_{{\rm out}}^{2}}{P_{{\rm in}}}(1-e_{{\rm out}}^{2})^{3/2}, \label{eq:tsec}
\end{equation}
where $m_{{\rm out}}$ is the mass of the outer companion, $m_{{\rm tot}}=m_{{\rm out}}+m_{{\rm bin}}$
is the total mass in the system, $m_{{\rm bin}}$ is the mass of the
inner binary, $e_{{\rm out}}$ is the outer eccentricity, $P_{{\rm in}}$
and $P_{{\rm out}}$ are the inner and outer orbital periods, respectively. 

The DA approximation neglects any osculating fluctuations of the orbital
elements on timescales $t\ll \tau_{{\rm sec}}$. However, in mildly hierarchical
systems, such shorter-term effects change the evolution of the triple,
and can induce larger eccentricites than predicted by the DA approach, as first shown by \citet{2012ApJ...757...27A}, while keeping an overall ``quasi-secular''
evolution (Lidov-Kozai cycles) very similar to that expected in the DA regime.
Accounting for the quasi-secular regime can be important for a wide
variety of systems at all scales \citep{cuk04,2012ApJ...757...27A,katz_sne,ant14,AMK14,Grishin17}. The rapid oscillations
identified near the maximal eccentricity have been considered in \citet{ant14,AMK14},
and recently, \citet{Katz16} have shown that the orbital elements
can be decomposed into averaged and fluctuating parts, and computed
the additional corrections due to single-averaged (SA) 
potential, providing consistent results with direct N-body integrations.

When the eccentricity (pericenter) is large (small), additional short-range
forces (e.g. tides, general-relativistic (GR) precession or tidal and rotational rotational
bulges) could affect and constrain the maximal eccentricity attained.
\citet{LML15} used conservation of the total potential energy and
$j_{z}$ to find the maximal eccentricity. For large enough strengths
of the extra forces, the eccentricity excitations can be suppressed
\citep{LML15}. 

Here we calculate the maximal eccentricity in the quadrupole order level of approximation and test particle limit, taking into account the
additional SA potential, the osculating oscillations of $j_{z}$
(and consequently in $e$) and the additional extra forces. Relaxing these limitations is discussion in sec. \ref{sec5}. We show
that contrary to quenching due to short range forces, the maximal
eccentricity is enhanced due to the dominating effect of fluctuations
in $j_{z}$. The enhancement may be orders of magnitude larger than
the widely used $e_{{\rm max}}^{{\rm DA}}$ (Eq. \ref{eq:emax}),
and even unconstrained, depending the level of the hierarchy and the
initial inclination. This, in turn have consequences for the mildly-hierarchical triples in all scales. Here we explore these effects and discuss their implications for two test cases - production of GW-sources near MBHs  and the formation and evolution of HJs. 

Our paper is organized as follows: In sec. \ref{sec2} we review basic LK mechanism
and its coupling to additional extra forces. In sec. \ref{sec3} we derive the
new formula for the maximal eccentricity in the quasi-secular CDA regime, and compare and validate
our results with N-body integrations. In sec. \ref{sec4}  We extend our analysis
to include extra forces. We apply our results to find the GW merger
time for Black-Hole binaries in the Galactic Centre (sec. \ref{41}), and then
compare the changes in the rate of HJ formation with the recent analytical
models (sec. \ref{42}). Finally, in sec. \ref{sec5} we discuss the limitations of
our model and summarize in sec. \ref{sec6}. 

\section{Coupling Lidov-Kozai with extra forces}\label{sec2}

The effects of the non-secular perturbations can effectively be considered as an additional perturbing extra-force or an effective corrected/perturbed potential. Including such pertubrations has been explored in the context of various non-Keplerian perturbations. It was used to derive the maximal eccentricity attained by the inner binary in a hierarchical triple, when affected by some given extra-forces. After a brief overview of the basic secular Lidov-Kozai approach, we describe the perturbative potential methods and its use in such contexts, such as finding the maximal eccentricity when accounting for general-relativistic precession and tidal effects. Equipped with these tools we follow a similar approach in exploring the non-secular SA effects and provide an analytic formulation for the maximal eccentricity in this regime. 
 
\subsection{Standard Lidov-Kozai potential} \label{21}
Consider an inner binary with masses $m_{0}$ and $m_{1}$ separated
by semimajor axis $a_{1}$ and eccentricity $e_{1}$, perturbed by
a companion of mass $m_{{\rm out}}$ and semimajor axis $a_{{\rm out}}$
and eccentricity $e_{{\rm out}}$. The DA quadrupole potential is
(e.g. \citealp{LML15}) is
\begin{equation}
\Phi_{{\rm quad}}=\frac{\Phi_{0}}{8}\left[1-6e_{1}^{2}-3(\boldsymbol{j}_{1}\cdot\hat{\boldsymbol{n}}_{2})^{2}+15(\boldsymbol{e}_{1}\cdot\hat{\boldsymbol{n}}_{2})^{2}\right],\label{eq:pot_quad}
\end{equation}
where $\Phi_{0}=Gm_{{\rm out}}m_{0}m_{1}a_{1}^{2}/(m_{{\rm bin}}a_{{\rm out}}^{3}(1-e_{{\rm out}}^{2})^{3/2})$
, $m_{{\rm bin}}=m_{0}+m_{1}$ is the binary mass, $\hat{\boldsymbol{n}}_{2}$
is the direction of the outer angular momentum, $\boldsymbol{e}_{1}=e_{1}\hat{\boldsymbol{e}}_{1}$
is the specific Laplace-Runge-Lenz (or eccentricity) vector, and $\boldsymbol{j}_{1}=\sqrt{1-e_{1}^{2}}\hat{\boldsymbol{j}}_{1}$
is the normalized angular momentum vector. 

Taking the reference frame $\hat{\boldsymbol{n}}_{2}=\hat{\boldsymbol{z}}$,
the $\boldsymbol{e}_{1},\boldsymbol{j}_{1}$ vectors can be expressed
in terms of the usual orbital elements (we drop the subscript $"1"$ for the
inner binary parameters for brevity):
\begin{align}
\boldsymbol{e} & =e\begin{pmatrix}\cos\omega\cos\Omega-\sin\omega\sin\Omega\cos i_{{\rm tot}}\\
\cos\omega\sin\Omega+\sin\omega\cos\Omega\cos i_{{\rm tot}}\\
\sin\omega\sin i_{{\rm tot}}
\end{pmatrix}\label{eq:evec}\\
\boldsymbol{j} & =\sqrt{1-e^{2}}\begin{pmatrix}\sin\Omega\sin i_{{\rm tot}}\\
-\cos\Omega\sin i_{{\rm tot}}\\
\cos i_{{\rm tot}}
\end{pmatrix},\label{eq:jvec}
\end{align}
where $\omega$ is the argument of pericenter, $\Omega$ is the argument
of ascending node and $i_{{\rm tot}}$ is the inclination angle between
the orbital planes of both binaries. Note that in the quadrupole approximation
in the test particle limit, $m_{1}\ll m_{{\rm out}}$,
the $\hat{\boldsymbol{z}}$ component on the inner angular monentum
is conserved, i.e. $j_{z}=\sqrt{1-e^{2}}\cos i_{{\rm tot}}={\rm const},$
and the outer angular momenta remaines fixed. Expressed
in orbital elements, the quadrupole potential is \citep{Naoz2016review}

\begin{equation}
\Phi_{{\rm quad}}=-\frac{\Phi_{0}}{8}\left[2+3e_{1}^{2}-3(1-e_{1}^{2}+5e_{1}^{2}\sin^{2}\omega_{1})\sin^{2}i_{{\rm tot}}\right].\label{eq:phi_quad2}
\end{equation}
The equations of motion can be solved for either
for the normalized vector pair ($\boldsymbol{e},\boldsymbol{j};$
\citealp{T9,LML15}) or for the orbital elements $(e,\omega,\Omega,i_{{\rm tot}};$
\citealp{Ford2000,naoz13gen}) equivalently. The maximal eccentricity
obtained is given by Eq. (\ref{eq:emax}).

\subsection{Non-Keplerian perturbations} \label{22}
 The standard LK mechanism is a property of purely
Newtonian point masses. In reality, additional non-Keplerian forces,
such as general relativistic (GR) corrections and tidal and rotational
bulges, may change the orbital evolution. The non-Keplerian extra
forces are strongest when the separation is smallest, thus these are effectively short-range forces. When the perturbation is weak, the forces are conservative,
and mainly cause extra precession of the apsidal angle $\omega$.
When the perturbation is strong, the typical dissipation timescales are short enough to change the orbital dynamics, and the forces are dissipative. The dissipation causes a 
loss of energy and angular momentum, circularizes the inner orbit and
brings it closer. Here we review the recent developments with connection
to the LK mechanism, focusing on the modified maximal eccentricity.

\subsubsection{General relativistic corrections} \label{221}

In order to take into account GR precession, the
leading order post-newtonian (PN) correction is \citep{b02,LML15,LL18}
\begin{equation}
\Phi_{{\rm GR}}=-\epsilon_{{\rm GR}}\Phi_{0}\frac{1}{(1-e_{1}^{2})^{1/2}}, \label{eq:phigr-1}
\end{equation}
 where 
\begin{equation}
\epsilon_{{\rm GR}}\equiv\frac{3m_{{\rm bin}}(1-e_{{\rm out}}^{2})^{3/2}}{m_{{\rm out}}}\left(\frac{a_{{\rm out}}}{a}\right)^{3}\frac{r_{g}}{a}\label{eq:epsgr-1}
\end{equation}
is the relative strength of GR precession. Here $r_{g}\equiv Gm_{{\rm bin}}/c^{2}$
is the gravitational radius. By comparing the (constant) total energy
$\Phi=\Phi_{{\rm quad}}+\Phi_{{\rm GR}}$ and $\hat{\boldsymbol{z}}$
angular momentum at two different locations of extremal eccentricites,
$e_{{\rm min}}\approx0$ and $e_{{\rm max}}$, \citep{LML15} found\footnote{See their Eq. (50) with $\epsilon_{{\rm Tide}}=\epsilon_{{\rm Rot}}=0$,
note they have a typo in the last term: $3\cos^{2}i_{0}/5$ should
be $5\cos^{2}i_{0}/3.$}
\begin{align}
\epsilon_{{\rm GR}}\left(\frac{1}{j_{{\rm min}}}-1\right) & =\frac{9}{8}\frac{e_{{\rm max}}^{2}}{j_{{\rm min}}^{2}}\left(j_{{\rm min}}^{2}-\frac{5}{3}\cos^{2}i_{0}\right), \label{eq:egr1}
\end{align}
where $j_{{\rm min}}\equiv\sqrt{1-e_{{\rm max}}^{2}}$ and $i_{0}$
is the initial inclination. For $\epsilon_{{\rm GR}}\ll1,$ GR precession
is slow compared to LK timescale. In this case the maximal eccentricity
is 

\begin{equation}
j_{{\rm min}}\approx\frac{4}{9}\epsilon_{{\rm GR}}\pm\frac{\sqrt{16\epsilon_{{\rm GR}}^{2}+135\cos^{2}i_{0}}}{9}. \label{eq:jmingrapprox-1}
\end{equation}
In the limit of of $\epsilon_{{\rm GR}}=0$ we get back to Eq. (\ref{eq:emax}). 

For ${\rm \epsilon_{{\rm GR}}\gg1},$ GR precession
is significant and the LK mechanism is quenched. For large enough $\epsilon_{{\rm GR}}$
the solutions approach $j_{{\rm min}}\to1$ (and $e_{{\rm max}}\to0$). 

If the bodies are too close, GW-wave induced dissipation
is important and the binary mill merge. The importance of GR corrections
is mostly relevant for compact object binaries and will be discussed
in the applications section.

\subsubsection{Tidal and rotational bulges} \label{222}

The additional potential raised by equilibrium tides
is \citep{2001ApJ...562.1012E,LML15}
\begin{equation}
\Phi_{{\rm Tide}}=-\epsilon_{{\rm Tide}}\frac{\Phi_{0}}{15}\frac{f_1(e_{1})}{(1-e_{1}^{2})^{9/2}}, \label{eq:phitides-1}
\end{equation}
 where $f_1(e)=1+3e^{2}+3e^{4}/8$ and 
\begin{equation}
\epsilon_{{\rm Tide}} \equiv \frac{15m_{0}^{2}a_{{\rm out}}^{3}(1-e_{{\rm out}}^{2})^{3/2}k_{2p,1}R_{1}^{5}}{a^{8}m_{1}m_{{\rm out}}}, \label{eq:epstide-1}
\end{equation}
where $k_{2p,1}$ is the Love number and $R_1$ is the radius  of body 1. Similarly to GR precession,
comparing the total potential $\Phi_{{\rm tot}}=\Phi_{{\rm quad}}+\Phi_{{\rm Tide}}$
for two extreme values of eccentricity yields the implicit equation
\citep{LML15}
\begin{align}
\frac{\epsilon_{{\rm Tide}}}{15}\left(\frac{f_1(e_{{\rm max}})}{8j_{{\rm min}}^{9}}-1\right) & =\frac{9e_{{\rm max}}^{2}}{8j_{{\rm min}}^{2}}\left(j_{{\rm min}}^{2}-\frac{5}{3}\cos^{2}i_{0}\right). \label{etide}
\end{align}
Similarly to the GR case, strong tidal bulges ($\epsilon_{{\rm Tide}}\gg1$)
will quench the LK oscilaltions and the binary will remain circular.
Note that for giant planets, the bulges are dominated by the planetary
oblateness, and the analysis is analogous (e.g. \citealp{T9,Grishin18}).

\subsubsection{Dissipative forces} \label{223}

When the two bodies are close, dissipation of energy
could be important. The typical timescale for dissipation for an isolated
binary of separation $a$ and eccentricity $e$ due to GW emission
is \citep{Peters64}
\begin{equation}
T_{{\rm m}}=\frac{5c^{5}a^{4}}{256G^{3}m_{1}m_{2}(m_{1}+m_{2})}(1-e^{2})^{7/2}. \label{eq:t_merge0-1}
\end{equation}
Usually this timescale is long even for tight compact object binary,
unless the eccentricity is large. LK oscillations can increase the
merger time \citep{Randall_LK1,Randall_LK2,LL18} and will be discussed
later.

For star-planet binaries, the migration time depends
on the (uncertain) internal structure of the planet and given by (Eq.
(9) of \citealp{hut81} and Eq. (26) of \citealp{Anderson_ps}) 
\begin{align}
\frac{1}{t_{a}} & \equiv\frac{1}{a}\frac{da}{at}=\frac{6k_{1}}{T}\frac{m_{1}}{m_{2}}\frac{m_{1}+m_{2}}{m_{2}}\left(\frac{R}{a}\right)^{8} \nonumber \\
 & = 6k_{1}n^{2}\tau_{L}\frac{m_{1}}{m_{2}}\left(\frac{R}{a}\right)^{5}, \label{eq:migration_time}
\end{align}
where $k_{1}$ is the apsidal motion constant, $\tau_{L}$ is the
tidal lag time and $T=R_{1}^{3}/(Gm_{2}\tau_{L})$ is the typical
time for changes in the orbit, and $n=\sqrt{G(m_{1}+m_{2})/a^{3}}$
is the mean motion. \footnote{For consistency with \citet{hut81} and \citet{Anderson_ps},
the apsidal motion constant $k$ is recognized as the tidal Love number
$k_{2p}$ in \citet{Anderson_ps}. For consistency with the definition
in Eq. (A9) of \citet{2007ApJ...669.1298F}, the viscous time is $t_{\nu1}=3(1+2k_{1})^{2}Tm_{2}/(2k_{1}m_{1})=3(1+2k_{1})^{2}R_{1}^{3}/(2k_{1}Gm_{1}\tau_{L})$.
Note that there are typos in footnote 2 of \citet{Pet15LKHJ}. For
typical values of $\tau_{L}=0.1\ {\rm s}$, $k_{1}=0.37$ and Jovian
parameters, the viscous time is $t_{\nu 1}\approx1\ {\rm yr}$. The $\sim1$
year viscous time is required for high-e migration. }

For coupled Kozai-Cycles and Tidal Friction (KCTF)
evolution, \citep{Anderson_ps} found that the dissipation time is
\begin{align}
\tau_{{\rm dis}} & =t_{a}\frac{(1-e_{{\rm max}}^{2})^{7}}{f_{2}(e_{{\rm max}})} \nonumber \\
f_{2}(e) & =1+\frac{31}{2}e^{2}+\frac{255}{8}e^{4}+\frac{185}{16}e^{6}+\frac{25}{64}e^{8}. \label{tau_dis_tides}
\end{align}

These dissipative effects are important when the
typical merger or dissipation timescale are comparable to the age of the system. The main difference is that tidal dissipation stops when the
orbit circularizes and synchronization of the orbit and the spin are
reached, while circular GW emitting binaries will continue to spiral
in until they merge.

\begin{figure*}
\begin{centering}
\includegraphics[width=18cm]{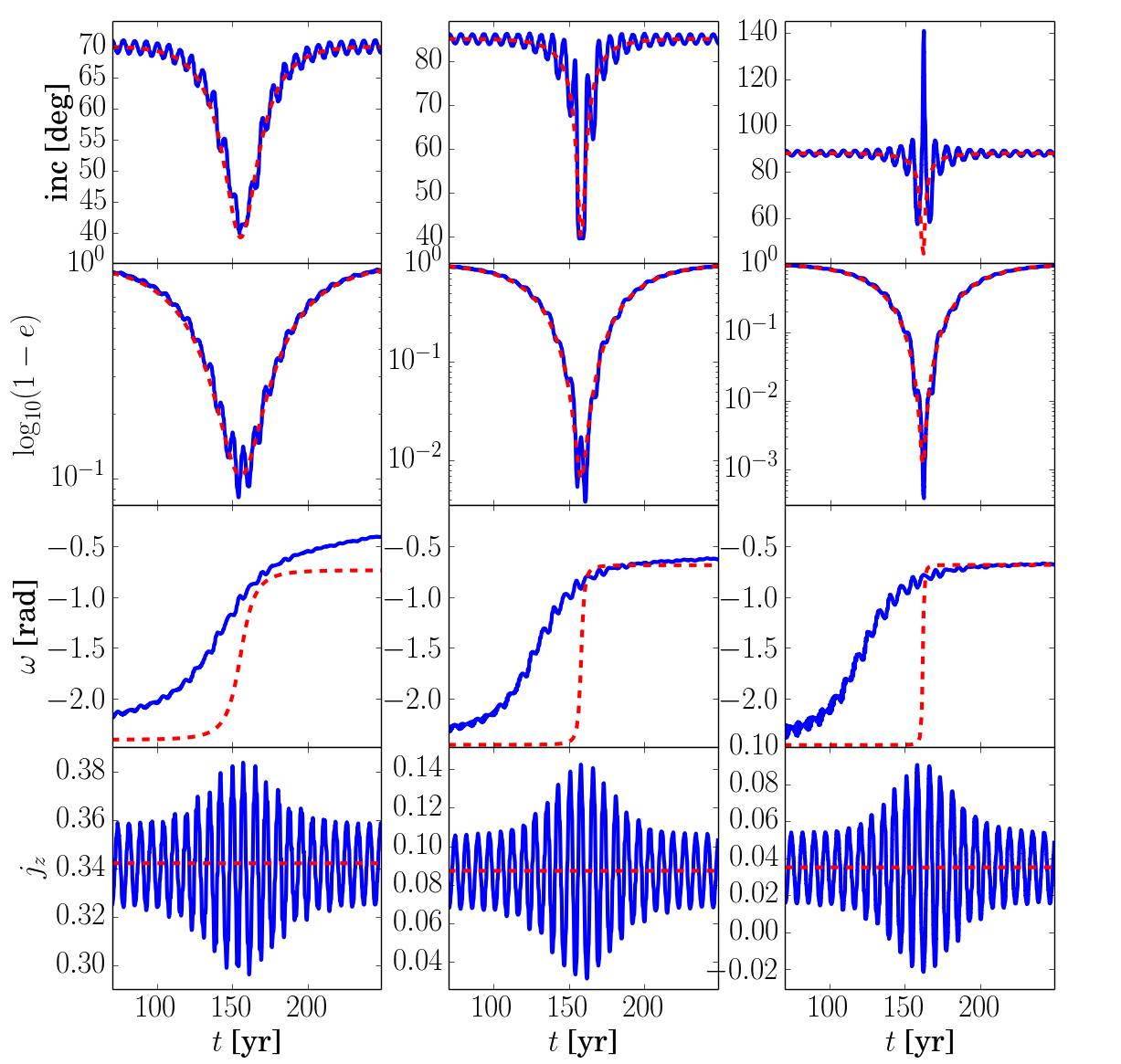}
\par\end{centering}
\caption{\label{fig:1}Example of a direct N-body integration (solid blue lines) vs double-averaged secular code (dashed red lines). The inner binary
has masses $M_{\odot}$ and $M_{J}$, $a_{1}=1{\rm \ AU},$ $e_{1}=0.001,$
$\Omega_{1}=\pi/4,$ $\omega_{1}=\pi/2$ and $f=0$. The outer mass
is $m_{{\rm out}}=3.2655\ M_{\odot}$, where the parameters of the
outer orbit are $a_{2}=10\ {\rm AU},$ $e_{2}=0.001,$ (corresponding
to $\epsilon_{{\rm SA}}=0.05$) $\Omega_{2}=\omega_{2}=f=0$. Top to bottom: Inclination, eccentricity, argument of pericentre, $j_z$. Left to right: Initial inclinations of $i_{{\rm tot}}=70,85,88^{\circ}$, corresponding to DA, SA and N-body regime, respectively.
We see that all the orbital
elements have a fast fluctuating term, modulated by the secular LK resonance. Even in the DA regime, $e_{\rm max}$ is under-predicted. The strong fluctuations in $j_{z}$ occur
near the maximal eccentricity. In N-body regime, the angular momentum flips sign and orbital flips are possible.  }
\end{figure*}

\subsection{Keplerian perturbations and corrected double averaging} \label{23}

The DA approximation misses perturbations on timescales shorter than
the secular timescale, $\tau_{{\rm sec}}$. When the hierarchy is
weak, the accumulated errors in neglecting these perturbations could
be large. Indeed, such effects have already been shown to be important
in various astrophysical systems \citep{cuk04,2012ApJ...757...27A,ant14}

It is possible to use the SA equations of motion that depend on the position of the outer orbit, $\boldsymbol{r}_{\rm out}$ \citep{Katz16,LL18}. \citet{Katz16} discuss the corrections to the double averaging approximation
from short-term oscillations. The key parameter that measures the
level of the hierarchy and the typical perturbations is the ``Single
Averaging'' (SA) strength ( Eq. 20 of \citealp{Katz16}) is
\begin{equation}
\epsilon_{{\rm SA}}\equiv\left(\frac{a_{1}}{a_{{\rm out}}(1-e_{{\rm out}}^{2})}\right)^{3/2}\left(\frac{m_{{\rm {\rm out}}}^{2}}{(m_{1}+m_{{\rm {\rm out}}})m_{1}}\right)^{1/2}=\frac{P_{{\rm out}}}{2\pi\tau_{{\rm sec}}}.\label{eq:epssa}
\end{equation}
 The vectors that describe the binary can be decomposed into the averaged
ones $(\bar{\boldsymbol{j}};\ \bar{\boldsymbol{e}})$ that vary slowly
on a secular timescale $\tau_{{\rm sec}}$ (Eq. \ref{eq:tsec}), and
the fluctuating ones ($\boldsymbol{j}_{f}\equiv\boldsymbol{j}-\bar{\boldsymbol{j}};\ \boldsymbol{e}_{f}\equiv\boldsymbol{e}-\bar{\boldsymbol{e}}$),
that vary with a period $\sim P_{{\rm out}}$ . The effects of a weak hierarchy are two-fold:
\begin{enumerate}
\item Short term fluctuations in the orbital elements with an amplitude
that depend on $\epsilon_{{\rm SA}}$ and the averaged values of
$(\bar{\boldsymbol{j}},\bar{\boldsymbol{e}})$:
\begin{equation}
\mathcal{A}(\epsilon_{{\rm SA}},\bar{\boldsymbol{j}},\bar{\boldsymbol{e}})=\epsilon_{{\rm SA}}\sqrt{C^{2}+S^{2}}\left(1+\frac{2\sqrt{2}}{3}e_{{\rm out}}\right),\label{eq:amp}
\end{equation}
where
\begin{align}
C & \equiv\frac{3}{8}\left(5\bar{e}_{x}^{2}-5\bar{e}_{y}^{2}-\bar{j}_{x}^{2}+\bar{j}_{y}^{2}\right)\nonumber \\
S & \equiv\frac{3}{8}\left(-10\bar{e}_{x}\bar{e}_{y}+2\bar{j}_{x}\bar{j}_{y}\right).\label{eq:cscs}
\end{align}

\item Additional evolution of the averaged vectors $(\bar{\boldsymbol{j}};\ \bar{\boldsymbol{e}})$
themselves. The full equations of motion appear in Appendix A of \citet{Katz16},
which corresponds to the additional effective ``corrected double
averaging'' potential (Eq. 39 of \citealp{Katz16}):
\begin{equation}
\Phi_{{\rm SA}}(\bar{\boldsymbol{j}},\bar{\boldsymbol{e}})=-\epsilon_{{\rm SA}}\frac{Gm_{{\rm out}}a_{1}^{2}}{a_{{\rm out}}^{3}(1-e_{{\rm out}}^{2})^{3/2}}\left(\phi_{{\rm circ}}+e_{{\rm out}}^{2}\phi_{{\rm ecc}}\right)\label{eq:phicda}
\end{equation}
where 
\begin{align}
\phi_{{\rm circ}}(\bar{\boldsymbol{j}},\bar{\boldsymbol{e}}) & =\frac{27}{64}\bar{j}_{z}\left\{ \frac{1-\bar{j}_{z}^{2}}{3}+8\bar{e}^{2}-5\bar{e}_{z}^{2}\right\}
\label{eq:phicdacirc} \\
\phi_{{\rm ecc}}(\bar{\boldsymbol{j}},\bar{\boldsymbol{e}}) & =\frac{3}{64}\left\{ \bar{e}_{z}(10\bar{j}_{x}\bar{e}_{x}-50\bar{j}_{y}\bar{e}_{y})\right.\nonumber \\
 & +\left.\bar{j}_{z}(5\bar{j}_{x}^{2}-\bar{j}_{y}^{2}+65\bar{e}_{x}^{2}+35\bar{e}_{y}^{2})\right\} 
\label{eq:phicdaecc}
\end{align}
\end{enumerate}
\citet{Katz16} have shown that adding these corrections is more compatible
with N-body integrations and it changes the long-term dynamics of particular orbits. Note that adding the corrected potential in Eq. (\ref{eq:phicda}) together with the fluctuating terms is equivalent to direct single-averaging (cf. Fig. 3 and 4 of \cite{Katz16} for comparison).

Previous studies have identified that DA is valid if $\sqrt{1-e_{\rm max}^2} \ge 2\pi\epsilon_{\rm SA}$, otherwise SA regime is valid if $\sqrt{1-e_{\rm max}^2} \ge 2\pi\epsilon_{\rm SA}^2$ \citep{LL18}. Thus, for eccentricities which exceed the latter limit, direct N-body integration is required (N-body regime). The corrected DA formalism alleviates the need to switch between SA and DA regimes, and both regimes are accounted for via the continuous parameter $\epsilon_{\rm SA}$. Nevertheless, the N-body regime is achieved only when the fluctuation in angular momentum is at least the order of itself \citep{ant17}. We show in sec. \ref{33} that our maximal eccentricity formula is valid wherever it is bound.

\section{Corrected maximal eccentricity}\label{sec3}

In this section we calculate the corrected maximal
eccentricity attained from the contributions of the single averaging.
The initial conditions for eliminating the fluctuating elements and
the typical strength of fluctuations is found in sec. \ref{31} The contribution
from the effective potential is calculated in \ref{32}, while the fluctuating
contribution is calculated in \ref{33} We then compare our result with
direct N-body realizations.

\subsection{Initial conditions and $j_{z}$ fluctuation} \label{31}

In order to compare the SA secular equation with N-body integrations
we need the initial conditions to match for the averaged vectors $(\bar{\boldsymbol{j}};\ \bar{\boldsymbol{e}})$.
To linear order in $\epsilon_{{\rm SA}}$, the fluctuating terms are
a (finite) sum of Fourier components where the lowest period is $P_{{\rm out}}$.
We focus on the key parameter $\bar{j}_{z}$. 

For zero outer eccentricity $e_{{\rm out}}=0$, the fluctuating term
is given by (e.g. Eq. (32) of \citealp{Katz16}) 
\begin{equation}
j_{z,f}\equiv j_{z}-\bar{j}_{z}=\epsilon_{{\rm SA}}\left(S\sin(2f_{2})-C\cos(2f_{2})\right), \label{eq:jf1}
\end{equation}
 where $C$ and $S$ are given in Eq. (\ref{eq:cscs}) and $f_{2}$
is the true anomaly of the outer orbit. For low initial eccentricity,
$e\ll1$, and using the definitions of $\boldsymbol{j}$ (Eq. \ref{eq:jvec})
we get
\begin{equation}
j_{z,f}=-\frac{3}{8}\epsilon_{{\rm SA}}\sin^{2}i_{{\rm tot}}\cos(2\delta) ,\label{eq:jf2}
\end{equation}
where $\delta\equiv\Omega_{1}-f_{2}$. Thus, if we want the initial
condition to correspond to the averaged $\bar{j}_{z}$ we need to
choose the initial angles such that $\delta=(\pi/4+n/2),$ $n=\{0,1,2...\}$.

The fluctuation in $j_{z}$ is maximal where the inner eccentricity
is the largest. From Eq. (\ref{eq:amp}), and for highly eccentric orbits,
the fluctuation amplitude for $j_{z}$, $\mathcal{A}(\epsilon_{{\rm SA}},\bar{j}\ll1,\bar{e}\to1)\equiv\Delta j_{z}$
is 

\begin{equation}
\Delta j_{z}=\frac{15}{8}\epsilon_{{\rm SA}}\bar{e}_{{\rm max}}^{2}\cos^{2}\bar{i}_{{\rm min}}=\frac{9}{8}\epsilon_{{\rm SA}}\bar{e}_{{\rm max}}^{2}, \label{eq:dj2}
\end{equation}
where we used the standard value of $\cos^{2}\bar{i}_{{\rm min}}=3/5$.
Since the correction is already of order $\mathcal{O}(\epsilon_{{\rm SA}})$,
any deviations caused by single-averaging will be at least $\mathcal{O}(\epsilon_{{\rm SA}})$,
thus corrections to Eq. (\ref{eq:dj2}) from different values of either
$\bar{e}_{{\rm max}}$ or $\cos\bar{i}_{{\rm min}}$ will be $\mathcal{O}(\epsilon_{{\rm SA}}^{2})$.

Fig. \ref{fig:1} shows the evolution of the orbital elements of a
typical triple system where direct N-body and secular equations of motions are compared \footnote{We move the time-series of the N-body results such that the time at $e_{\rm max}$ will coincide.}. We see that the fluctuations in $j_{z}$ are
strongest where the eccentricity approaches $e_{{\rm max}}.$ In addition,
choosing $\delta=\pi/4$ guarantees that both $i_{{\rm tot}}$ and
$j_{z}$ will be set at their mean values. The actual maximal eccentricity
is larger than its averaged value. The eccentricity and $j_z$ panels show us that regardless of initial $j_z$ and the typical regime (DA, SA or N-body), the eccentricity is always underestimated with similar amplitude. Moreover, if $\Delta j_z > j_z$, orbital flips are allowed and the eccentricity is stochastic and unbound. In the next sections we calculate
the maximal eccentricity taking into account the short-term fluctuations
and compare the results with full N-body simulations.

\begin{figure*}
\begin{centering}
\includegraphics[width=8.2cm]{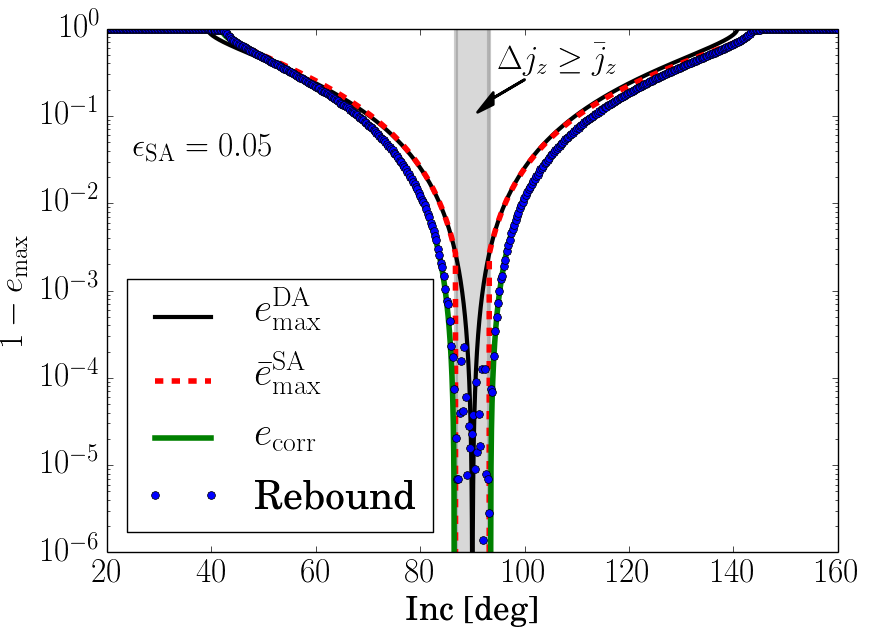}
\includegraphics[width=8.2cm]{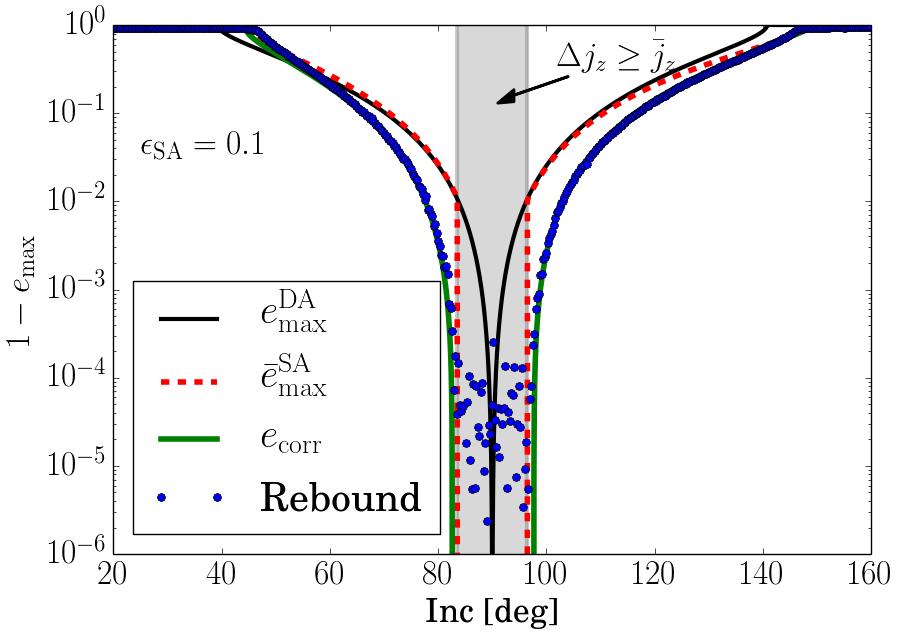}
\par\end{centering}
\begin{centering}
\includegraphics[width=8.cm]{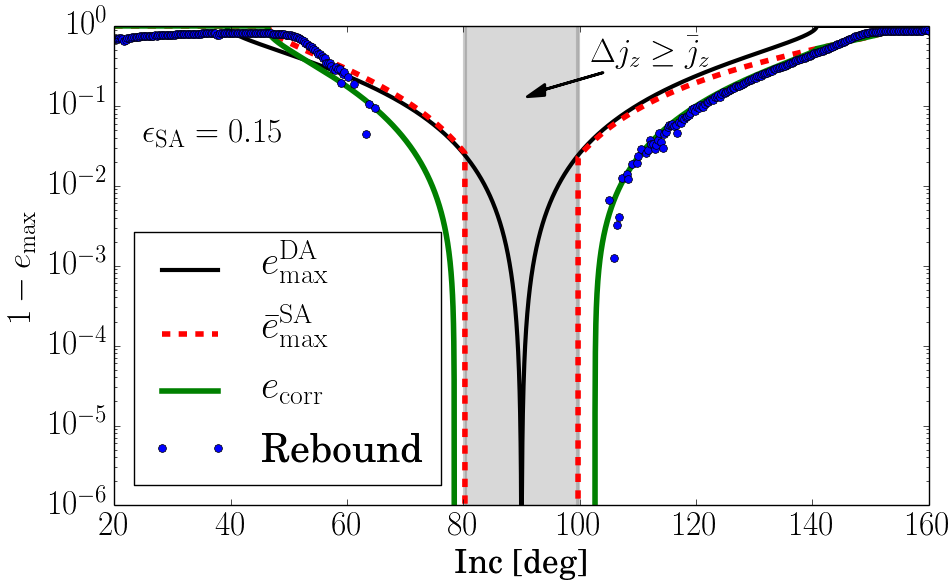}
\includegraphics[width=8cm]{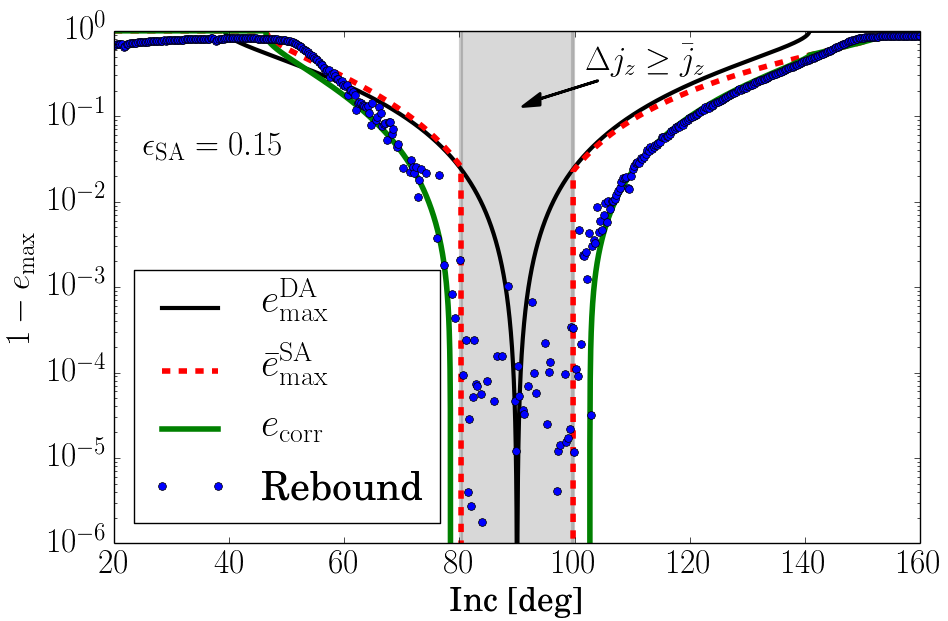}
\par\end{centering}
\caption{\label{fig:nbody}Maximal eccentricity versus initial inclination.
We compare corrected averaged secular theory vs. N-body realizations.
All simulations start with inner binary of masses $M_{\odot}$ and
$M_{J}$, $a_{1}=1{\rm \ AU},$ $e_{1}=0.001,$ $\Omega_{1}=\pi/4,$
$\omega_{1}=\pi/2$ and $f_1=0$. The outer orbit is at separation $a_{2}=10\ {\rm AU},$
$e_{2}=0.001,$ $\Omega_{2}=\omega_{2}=f_2=0$. Top left: $m_{{\rm out}}=3.2655\ M_{\odot}$
(corresponding to $\epsilon_{{\rm SA}}=0.05$). Top right: $m_{{\rm out}}=10.93\ M_{\odot}$
(corresponding to $\epsilon_{{\rm SA}}=0.1$). Bottom: $m_{{\rm out}}=23.461\ M_{\odot}$
(corresponding to $\epsilon_{{\rm SA}}=0.15$). The end time of all
runs is $10^{4}$ times the inner orbit, expect for top right, with
end time of $500$ inner orbits. Solid black line is the classical
$e_{\rm max}^{\rm DA}$ from standard double-averaged LK mechanism.
Dashed red line is the single-averaged corrected eccentricity $\bar{e}_{{\rm max}}^{{\rm SA}}$
given in Eq. (\ref{eq:emaxsa}). Solid green line is the corrected maximal
eccentricity after taking into account fluctuating terms in $e_{{\rm corr}}$
given in Eq. (\ref{eq:ecorr}). Grey area is the allowed zone for orbital
flips, where $1-e_{{\rm max}}$ is unbound. For $\epsilon_{{\rm SA}}=0.15$,
the system is very close to its Hill stability radius ($a_{1}\approx0.41r_{\rm H},$
$r_{\rm H}=10(M_{{\rm in}}/3M_{{\rm out}})^{1/3}$), hence highly inclined
orbits are unstable on long timescales \protect\citep{Grishin17}. Bottom left panel
shows integrating for shorter times and hence more orbits, which become
unstable after longer integrations. }
\end{figure*}

\subsection{Maximal single-averaged eccentricity} \label{32}

We are interested in finding $\bar{j}_{{\rm min}}(\bar{j}_{z},\bar{e})\equiv(1-\bar{e}_{{\rm max}}^{2})^{1/2}$
(and therefore $\bar{e}_{{\rm max}}=(1-\bar{j}_{{\rm min}}^{2})^{1/2}$
) as a function of initial conditions. We obtain it from equating
the total potential 
\begin{equation}
\Phi_{{\rm tot}}(\bar{\boldsymbol{j}},\bar{\boldsymbol{e}})=\Phi_{{\rm quad}}(\bar{j}_{z},\bar{e},\bar{e}_{z})+\epsilon_{{\rm SA}}\Phi_{{\rm SA}}(\bar{\boldsymbol{j}},\bar{\boldsymbol{e}})\label{eq:phitot}
\end{equation}
 in two points of extreme (minimal and maximal) eccentricities. A
similar approach to calculate $j_{{\rm min}}$ in the presence of
non-Keplerian forces was used in \citet{LML15} and reviewed in sec. \ref{22} 

To first order for circular orbits, $\bar{j}_{z}$ is conserved \citep{Katz16}
and the (dimensionless) potential depends on 
\begin{equation}
\phi_{{\rm tot}}(\bar{\boldsymbol{j}},\bar{\boldsymbol{e}})=\phi_{{\rm quad}}(\bar{j}_{z},\bar{e},\bar{e}_{z})+\epsilon_{{\rm SA}}\phi_{{\rm circ}}(\bar{j}_{z},\bar{e},\bar{e}_{z}).\label{eq:phitotdimless}
\end{equation}
 where $\phi_{{\rm tot}}\equiv\Phi_{{\rm tot}}/\Phi_{0}$. Note that for $e_{{\rm out}}\ne0$ orbits, $\bar{j}_{z}$ is no longer
a constant and we cannot close the equation to obtain the maximal
eccentricity\footnote{However, see \citet{Katz11} for an additional constant
of motion and analytic flip criteria. Finding the maximal eccentricity
where the outer perturber is eccentric is beyond the scope of this
paper.}. The maximal eccentricity from the additional SA evaluation term
is $\bar{e}_{{\rm max}}^{{\rm SA}}.$ Denote the
initial mutual inclination by $i_{{\rm tot}}=i_{0}$ and the inner
eccentricity is $e_{0}$ . In order to evaluate the potential in Eq.
(\ref{eq:phitotdimless}) we need to specify $\omega.$ For librating
orbits, $\omega=\pi/2$ for both extreme value of the eccentricity
\citep{Katz11}. For small minimal eccentricity, i.e.  $e_{0}\ll1$
the orbit could be circulating, but $\omega$ is not properly defined
and plays a role, since the term that contains $\omega$ is
proportional to $e^{2}$. Thus, it is safe to take $\omega=\pi/2$
in our evaluations, similarly to \citet{LML15}. 

For the initial conditions stated above, the potential is
\begin{align}
\phi_{{\rm tot}}(e_{0}) & =1+9e_{0}^{2}-3\bar{j}_{z}^{2}-15e_{0}^{2}\cos^{2}i_{0}\nonumber \\
 & -\epsilon_{{\rm SA}}\frac{27}{8}\bar{j}_{z}\left(\frac{1-\bar{j}_{z}^{2}}{3}+3e_{0}^{2}+5e_{0}^{2}\cos^{2}i_{0}\right). \label{eq:phitote2}
\end{align}

When the orbit attains its maximal eccentricity, the orbital elements are $e=\bar{e}_{{\rm max}},$
$\cos^{2}i_{{\rm min}}=\bar{j}_{z}^{2}/\bar{j}_{{\rm min}}^{2}$ and
$\omega=\pi/2$. The potential is 
\begin{align}
\phi_{{\rm tot}}(\bar{e}_{{\rm max}}) & =1+9\bar{e}_{{\rm max}}^{2}-3\bar{j}_{z}^{2}-15\bar{e}_{{\rm max}}^{2}\frac{\bar{j}_{z}^{2}}{\bar{j}_{{\rm min}}^{2}}\nonumber \\
 & -\frac{27}{8}\epsilon_{{\rm SA}}\bar{j}_{z}\left(\frac{1-\bar{j}_{z}^{2}}{3}+3\bar{e}_{{\rm max}}^{2}+5\bar{e}_{{\rm max}}^{2}\frac{\bar{j}_{z}^{2}}{\bar{j}_{{\rm min}}^{2}}\right). \label{eq:phitote1}
\end{align}

Equating both terms (Eq. \ref{eq:phitote1} and \ref{eq:phitote2})
we get 
\begin{align}
\bar{j}_{{\rm min}}^{2}\left(1-\frac{e_{0}^{2}}{\bar{e}_{{\rm max}}^{2}}\right) & =\frac{5}{3}\bar{j}_{z}^{2}\left(1-\frac{j_{{\rm min}}^{2}}{\bar{e}_{{\rm max}}^{2}}\frac{e_{0}^{2}}{j_{{\rm 0}}^{2}}\right)\nonumber \\
  +\frac{9}{8}\epsilon_{{\rm SA}} & \bar{j}_{z}\left[\bar{j}_{{\rm min}}^{2}\left[1-\frac{1}{3}\frac{e_{0}^{2}}{\bar{e}_{{\rm max}}^{2}}\left(3+5\frac{\bar{j}_{z}^{2}}{j_{{\rm 0}}^{2}}\right)\right]+\frac{5}{3}\bar{j}_{z}^{2}\right]. \label{eq:comp1}
\end{align}
In the limit of $e_{0}\ll1,$ $j_{0}\to1$ and $\bar{j}_{z}\to\cos i_{0}$ we
have 
\begin{equation}
\bar{j}_{{\rm min}}^{2}-\frac{5}{3}\cos^{2}i_{0}=\frac{9}{8}\epsilon_{{\rm SA}}\cos i_{0}\left(j_{{\rm min}}^{2}+\frac{5}{3}\cos^{2}i_{0}\right), \label{eq:comp2}
\end{equation}
or solving for $\bar{j}_{\rm min}$:
\begin{align}
\bar{j}_{{\rm min}}^{2} & =\frac{5}{3}\cos^{2}i_{0}\frac{1+\frac{9}{8}\epsilon_{{\rm SA}}\cos i_{0}}{1-\frac{9}{8}\epsilon_{{\rm SA}}\cos i_{0}}\nonumber \\
 & \approx\frac{5}{3}\cos^{2}i_{0}\left(1+\frac{9}{4}\epsilon_{{\rm SA}}\cos i_{0}\right)+\mathcal{O}(\epsilon_{{\rm SA}}^{2}).\label{eq:jzmin}
\end{align}
Note that the (averaged) maximal eccentricity in the SA regime is:
\begin{equation}
\boxed{\bar{e}_{{\rm max}}^{{\rm SA}}=\sqrt{1-\frac{5}{3}\cos^{2}i_{0}\frac{1+\frac{9}{8}\epsilon_{{\rm SA}}\cos i_{0}}{1-\frac{9}{8}\epsilon_{{\rm SA}}\cos i_{0}}}}.\label{eq:emaxsa}
\end{equation}
For linear terms in $\epsilon_{{\rm SA}}$
\begin{align}
\bar{e}_{{\rm max}}^{{\rm SA}} & \approx\sqrt{1-\frac{5}{3}\cos^{2}i_{0}\left(1+\frac{9}{4}\epsilon_{{\rm SA}}\cos i_{0}\right)}\nonumber \\
 & =\bar{e}_{{\rm max}}^{{\rm DA}}-\frac{15}{8\bar{e}_{{\rm max}}^{{\rm DA}}}\cos^{3}i_{0}\epsilon_{{\rm SA}}+\mathcal{O}(\epsilon_{{\rm SA}}^{2}).\label{eq:emaxsalin}
\end{align}
where the standard DA eccentricity $\bar{e}_{{\rm max}}^{{\rm DA}}$
is defined in Eq. (\ref{eq:emax}).

The critical inclination for the onset of the LK mechanism is obtained
in Appendix A
\begin{equation}
\cos i_{{\rm crit}}=\sqrt{\frac{3}{5}}-\frac{27}{40}\epsilon_{{\rm SA}}.\label{eq:inc_crit}
\end{equation}

Note that $\bar{e}_{{\rm max}}^{{\rm SA}}$ and $i_{{\rm crit}}$
break the symmetry between prograde and retrograde orbits, since $\cos i_{{\rm crit}}$
is no longer symmetric, which has implications on the general evolution
and Hill-stability of the system (\citealp{Grishin17}, Appendix A).

\begin{figure*}
\begin{centering}
\includegraphics[width=18cm]{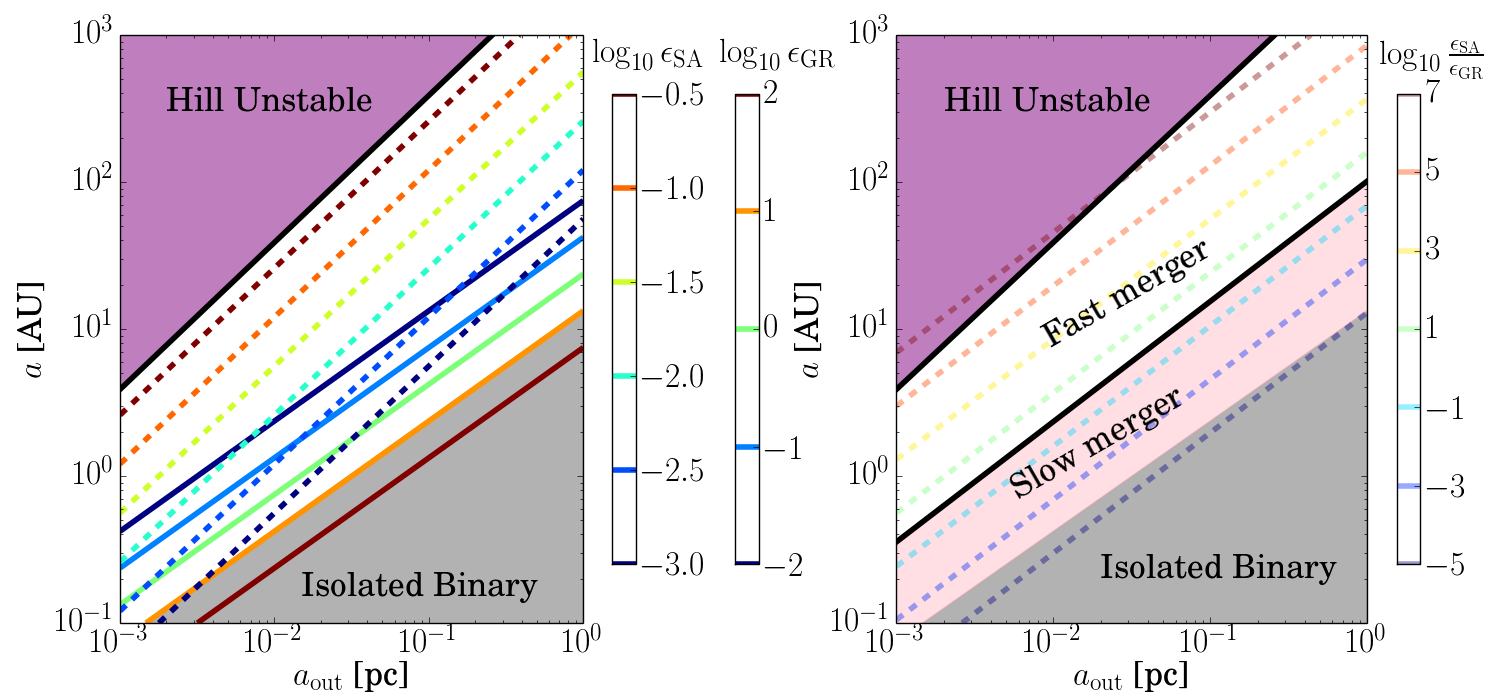}
\par\end{centering}
\caption{\label{fig:epsilons}Dimensionless parameters of BH binary as a function
of inner and outer separations. Left: Numerical values of $\epsilon_{{\rm GR}}$
(solid) and $\epsilon_{{\rm SA}}$ (dashed) on the $a$, $a_{{\rm out}}$
plane. Purple area (upper left corner) is the region where the binary
is Hill unstable, e.g. where $a>r_{{\rm H}}$ where $r_{{\rm H}}=a_{{\rm out}}(m_{{\rm bin}}/3m_{{\rm out}})^{1/3}$
is the Hill radius. Grey area (bottom right corner) is the region
where $\epsilon_{{\rm GR}}\ge10,$ so GR completely quenches eccentricity
excitations and binary evolves as an isolated binary. Right: Contours
of the ratio $\epsilon_{{\rm SA}}/\epsilon_{{\rm GR}}$. Pink area
is the region where $\epsilon_{{\rm GR}}\gg\epsilon_{{\rm SA}}$,
thus the maximal eccentricity is never close to unity, so only slow
mergers are possible. While area is where $\epsilon_{{\rm SA}}\gg\epsilon_{{\rm GR}},$
thus arbitrary large eccentricities and orbital flips are possible
if $\bar{j}_{z}$ is small enough. In this area fast merger (or direct
BH-BH collisons) are possible. The dividing line between slow and fast
mergers is given by Eq. (\ref{eq:flipgr}). }
\end{figure*}

\subsection{Maximal fluctuating eccentricity } \label{33}

We calculated $\bar{e}_{{\rm max}}^{{\rm SA}}$ from equating the
potential at two points with a constant $\bar{j}_{z}$. In reality,
$j_{z}$ fluctuates around an averaged value $\bar{j}_{z}$ and a
fluctuating amplitude $\Delta j_{z}$ given in Eq. (\ref{eq:dj2}).
In turn, the eccentricity is also fluctuating around an averaged value
$\bar{e}_{{\rm max}}$ and some fluctuation $\delta e$. Since $\phi_{{\rm circ}}\propto\bar{j}_{z}$,
when $\bar{j}_{z}\approx\cos i_{{\rm tot}}\ll1$ is small, the additional
term is of order $\mathcal{O}(\bar{j}_{z}\epsilon_{{\rm SA}}),$ therefore
we need to take into account second order terms in the expansion of
$\delta e_{{\rm max}}$, i.e. keeping terms of order $\mathcal{O}(\epsilon_{{\rm SA}}^{2},\bar{j}_{z}\epsilon_{{\rm SA}})$.

The corrected maximal eccentricity is 
\begin{align}
e_{{\rm corr}} & =\bar{e}_{{\rm max}}^{{\rm SA}}+\delta e=\sqrt{1-(\bar{j}_{{\rm min}}-\delta j)^{2}}\nonumber \\
\delta j & =\frac{\Delta j_{z}}{\cos i_{{\rm min}}}=\frac{9}{8}\sqrt{\frac{5}{3}}\left(\bar{e}_{{\rm max}}^{{\rm SA}}\right)^{2}\epsilon_{{\rm SA}}. \label{eq:ecorr}
\end{align}
 Since $j_{{\rm min}}\sim\mathcal{O}(\epsilon_{{\rm SA}})$, the leading
term is $\delta j\cdot j_{{\rm min}}\sim\mathcal{O}(\epsilon_{{\rm SA}}^{2})$,
therefore we need to use a Taylor expansion to second order: 
\begin{align}
|\delta e| & =\left.\frac{\partial e_{{\rm max}}}{\partial j}\right|_{j_{{\rm min}}}\delta j_{{\rm min}}+\frac{1}{2}\left.\frac{\partial^{2}e_{{\rm max}}}{\partial j^{2}}\right|_{j_{{\rm min}}}\delta j_{{\rm min}}^{2}\nonumber \\
 & =\frac{|j_{{\rm min}}|}{e_{{\rm max}}}\delta j_{{\rm min}}+\frac{(\delta j_{{\rm min}})^{2}}{2e_{{\rm max}}^{3}}. \label{eq:de1}
\end{align}
Plugging $j_{{\rm min}}=\bar{j}_{{\rm min}}-\delta j$ and $e_{{\rm max}}=\bar{e}_{{\rm max}}^{{\rm SA}}$
from Eq. (\ref{eq:emaxsa}) and (\ref{eq:ecorr}) yields

\begin{equation}
\boxed{\delta e=\frac{135}{128}\bar{e}_{{\rm max}}^{{\rm SA}}\epsilon_{{\rm SA}}\left\{ \frac{16}{9}\sqrt{\frac{3}{5}|}\bar{j}_{{\rm min}}|+\epsilon_{{\rm SA}}-2\epsilon_{{\rm SA}}\left(\bar{e}_{{\rm max}}^{{\rm SA}}\right)^{2}\right\} }. \label{eq:de3}
\end{equation}
where the absolute value of $|j_{{\rm min}}|$ accounts for retrograde
orbits. 

Note that \citet{ant14} also obtained
an expression for the fluctuation in the maximal eccentricity (their
Eq. 3). \citet{ant14} used the fluctuation of the orbit's
angular momentum near the maximal eccentricity, previously derived
in \protect{\citealp{Ivanov05}};  last Eq. B14). \citet{ant14} have taken incorrect mass dependence and prefactors in their Eq. (3). In appendix B we re-derive
the eccentricity fluctuation from the \protect{\citet{Ivanov05}} formula and
compare to our results. Our new formula, Eq. (\ref{eq:de3}) thus has three
new ingredients: the dependence on $e_{{\rm max}}$, namely $\delta e\propto e_{{\rm max}}$,
the use of $\bar{e}_{{\rm max}}^{{\rm SA}}$ instead of $\bar{e}_{{\rm max}}^{{\rm DA}}$,
and most important is the last term, which is proportional to $\propto\epsilon_{{\rm SA}}\bar{e}_{{\rm max}}^{2}$.
We show that \citet{ant14} overestimate the actual fluctuation, while
the error is increasing with increasing $\epsilon_{{\rm SA}}$ (see
appendix B for details). 

Figure \ref{fig:nbody} shows the comparison of the various prescriptions
for the maximal eccentricity with direct N-body integrations. For
N-body integrations we use the publicly available code \texttt{REBOUND} \citep{ReboundMain}.
We use \texttt{IAS15}, a fast, adaptive, high-order integrator for gravitational
dynamics, accurate to machine precision over a billion orbits \citep{ReboundIAS15}.
Overall, the simulation tends to follow the curve of $e_{{\rm corr}}$
(Eq. \ref{eq:ecorr}) For various values of $\epsilon_{{\rm SA}}$.
Note that in the region $|\bar{j}_{z}|\le\Delta j_{z}$, the value
of $1-e_{{\rm max}}$ is unbound from below. In this regime, the orbital orientation can flip from prograde to retrograde and vise versa, similarly to the orbital flip in the octupole regime (see \citealp{Naoz2016review} and discussion in sec. \ref{sec5}). 

One may still be cautious about the validity of Eq. (\ref{eq:de3}). The SA regime breaks down when $\sqrt{1-e_{\rm max}^2} \le 2\pi\epsilon_{\rm SA}^2$, since the time spent near $e_{\rm max}$ is shorter than $P_{\rm in}$. However, the eccentricity is bound only if 
\begin{equation}
\bar{j}_z = \sqrt{\frac{3}{5}(1-e_{\rm max}^2)} \ge \Delta j_z = \frac{9}{8} \epsilon_{\rm SA}. \label{nb}
\end{equation}
Thus, the SA equations break down and the eccentricity is bound only if $\epsilon_{\rm SA} > 9/(16\pi)\sqrt{5/3}\approx 0.23$.
Typical systems are almost always dynamically unstable for such large values of $\epsilon_{\rm SA}$. Thus, in most cases the flip criteria will be satisfied and the eccentricity will be unbound much before the SA equations will breakdown. 

To summarize, the corrected maximal eccentricity given in Eq. (\ref{eq:ecorr})
is the sum of two new terms: the averaged SA value given in Eq. (\ref{eq:emaxsa})
and the extra fluctuating value given in (\ref{eq:de3}) evaluated
at the minimum of the instantaneous value of $|j_{z}|$. The formula
is in excellent correspondence with direct N-body integrations for
all the examined parameters. In the limit $e_{{\rm {\rm max}}}\to1$,
the flip criteria $|\bar{j}_{z}|\le\Delta j_{z}$ is restored.

\section{Applications}\label{sec4} 

\begin{figure*}
\begin{centering}
\includegraphics[width=8cm]{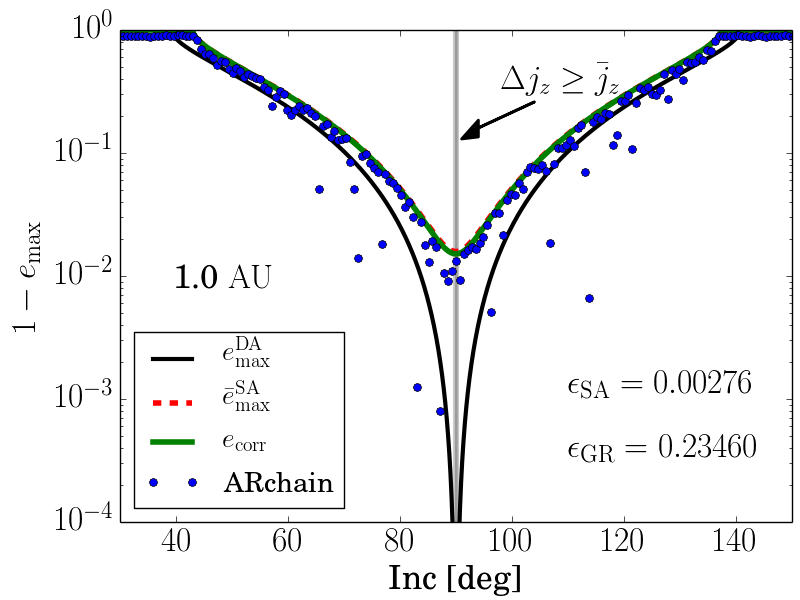}\includegraphics[width=8cm]{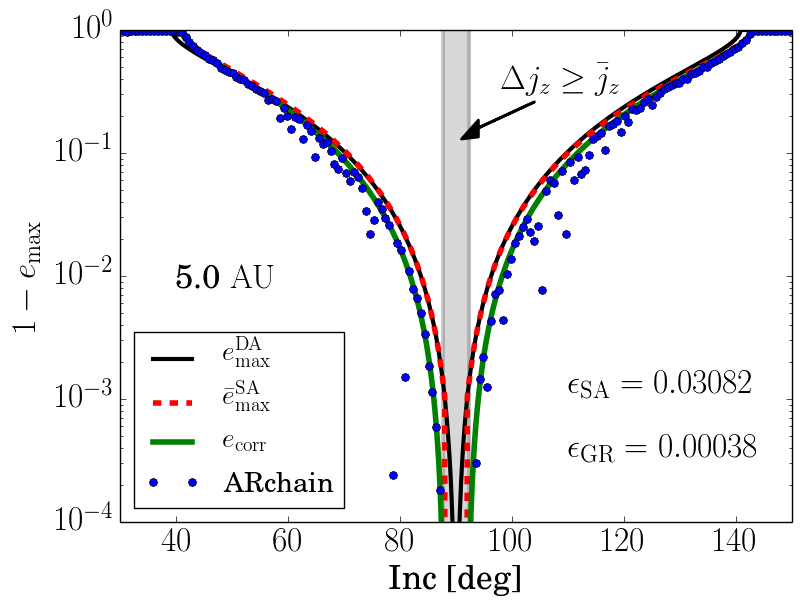}
\par\end{centering}
\begin{centering}
\includegraphics[width=8cm]{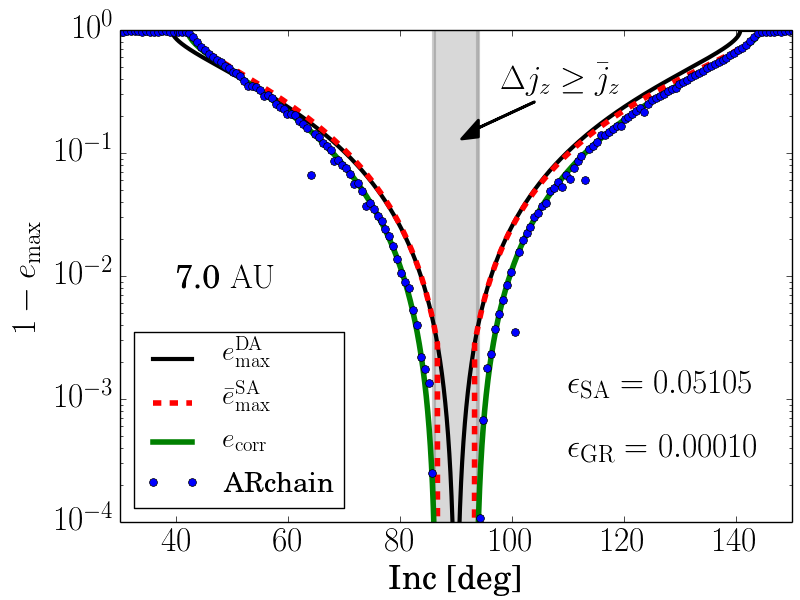}\includegraphics[width=8cm]{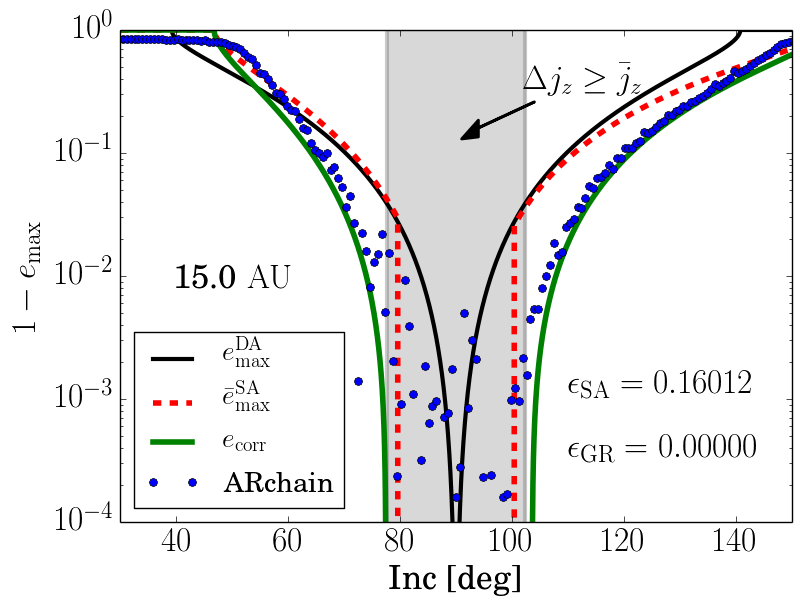}
\par\end{centering}
\caption{\label{fig:archain} Comparison of analytical estimates with direct
integration. We use \texttt{ARCHAIN} code. The initial conditions are $m_{1}=m_{2}=30M_{\odot}$,
$m_{3}=4\cdot10^{6}M_{\odot},$ $a_{2}=0.01\ {\rm pc},$ $\omega_{1}=\pi/2,$
$\Omega_{1}=\pi/4$, $e_{1}=e_{2}=0$. The other angles are zero.
Each panel shows the maximal eccentricity as a function of the initial inclination.
Each panel has 200 different realization of initial inclination $i_{0}\in[20,160]$
(blue dots), compared with our analytical prediction (green line,
Eq. \ref{eq:de3}, \ref{eq:jmin_e0}). The initial separation is indicated
on the left, with calculated valeus of $\epsilon_{{\rm SA}},$ $\epsilon_{{\rm GR}}$
on the right. Top row: $1,5\ {\rm AU},$ bottom row: $7,15\ {\rm AU}.$}
\end{figure*}

\begin{figure*}
\begin{centering}
\includegraphics[height=5.5cm]{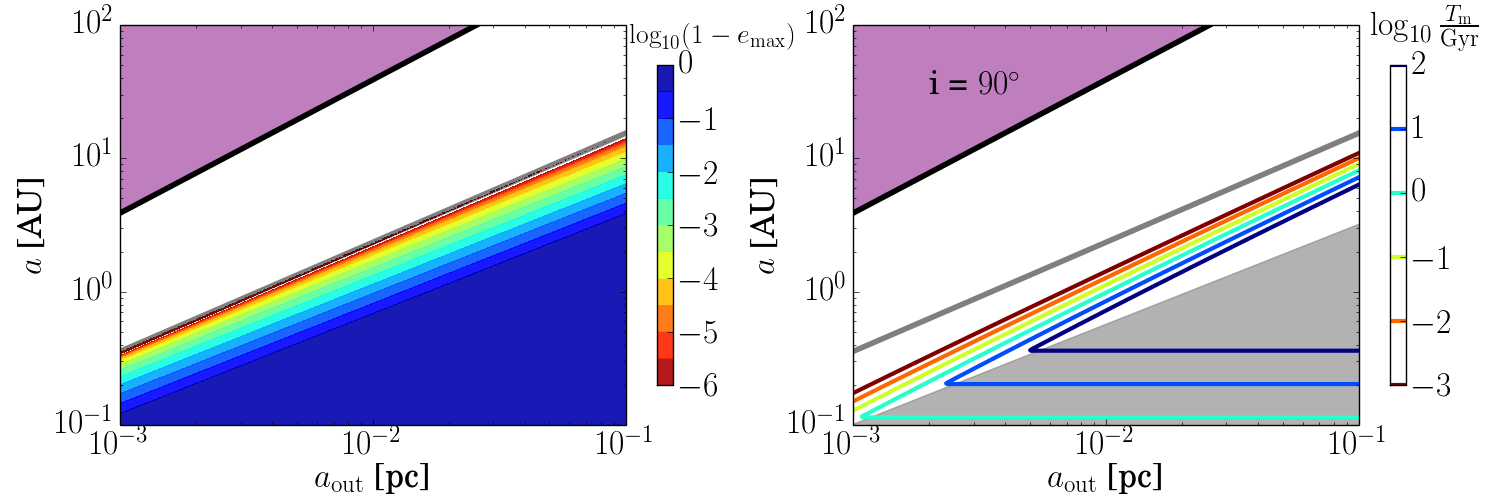}
\par\end{centering}
\begin{centering}
\includegraphics[height=5.5cm]{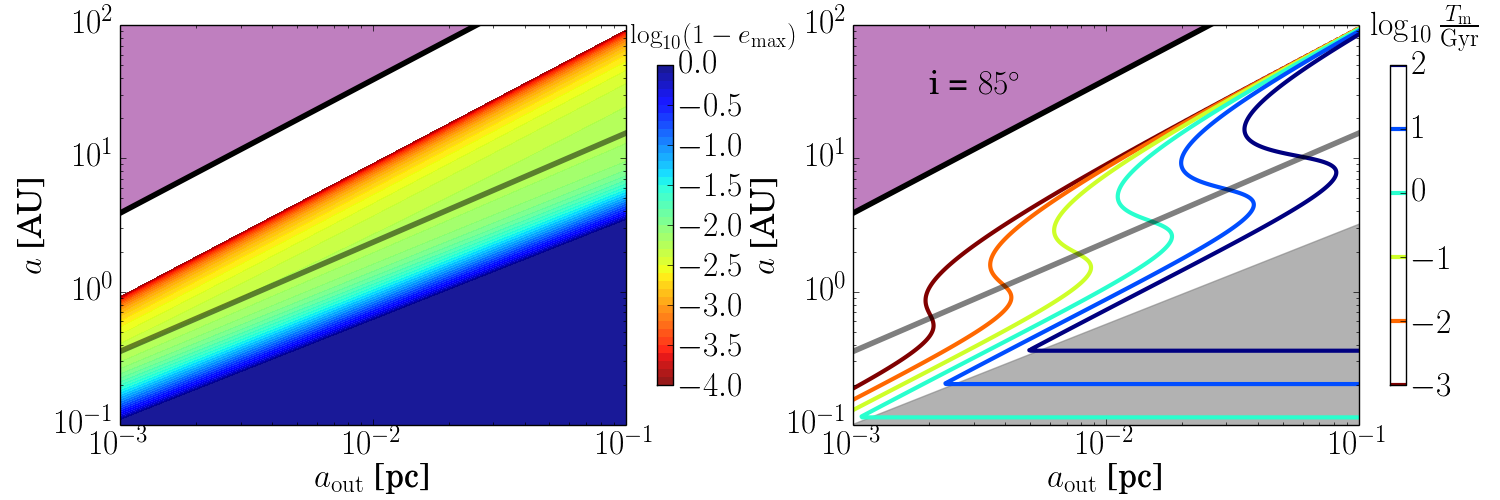}
\par\end{centering}
\begin{centering}
\includegraphics[height=5.5cm]{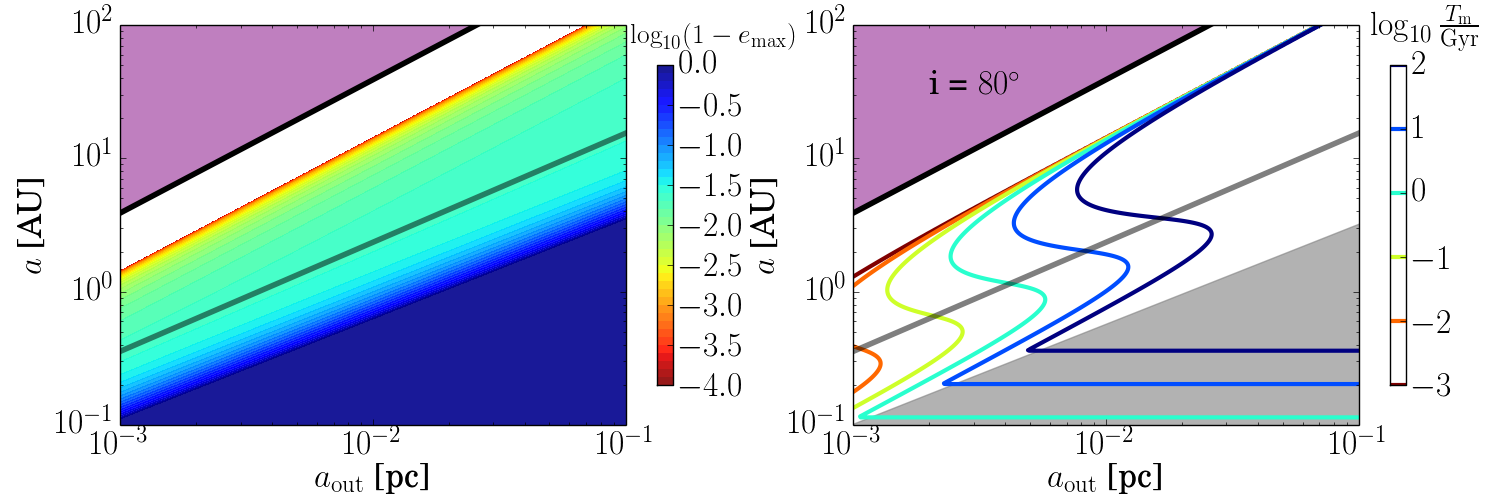}
\par\end{centering}
\begin{centering}
\includegraphics[height=5.5cm]{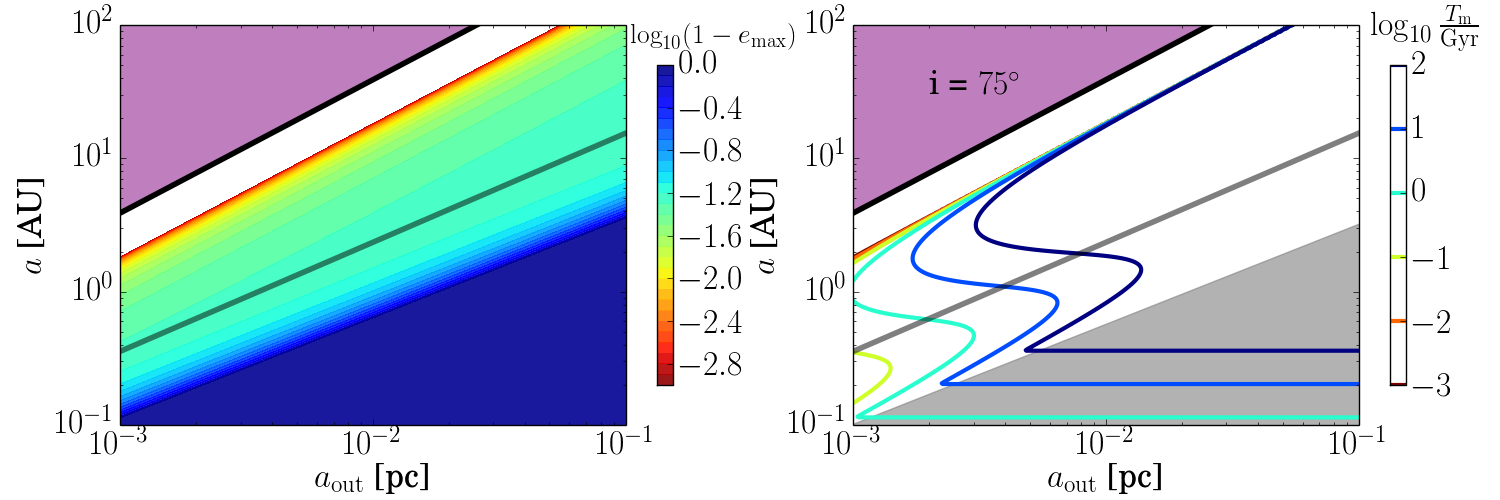}
\par\end{centering}
\caption{\label{fig:merge_times}Maximal eccentricy and merger times for binaries
in the Galactic Centre. Left panels: Maximal eccentricity as a function
of inner and outer semi-major axes. The transparent grey line is Eq.
(\ref{eq:flipgr}), which divides the areas where $e_{{\rm max}}$ is bound
and unbound at $i=90^{\circ}.$ Right: Merger times derived from Eq.
(\ref{eq:tmerge}). Grey and purple areas similar to Fig. \ref{fig:epsilons}
(i.e. Regions of isolated binary and Hill Unstable orbits, respectively).
Top to bottom: inclinations of $90,85,80,75$ degrees, respectively.}
\end{figure*}

\subsection{General relativistic corrections and GW mergers} \label{41}

The recent discoveries of high rates of gravitational-Wave (GW) mergers
of stellar black holes ($40-213\ {\rm Gpc}^{-3}{\rm yr^{-1}}$,
\citealp{abb17,abbott16,abbott17}) raise the question of their astrophysical
origin, and various possibilities exist. These include isolated binary
evolution \citep{bel16}, dynamically formed/evolved binaries in dense
stellar systems \citep[e.g.][and references therein]{askar17,rod18, FK18},
triple secular evolution of stellar binaries orbiting massive black
holes (MBHs; \citealp{2012ApJ...757...27A}), stellar triple systems
\citep{ant17}, and gas-assisted mergers near massive black holes \citep{bart17, metzger17}. 

One way to decrease the merger time and increase the merger rate is
pumping the eccentricity of the inner binary due to LK oscillations.
The resulting actual merger time is \citep{Randall_LK2,Randall_LK1,LL18}
\begin{equation}
T_{{\rm m}}=T_{m,0}(1-e_{{\rm max}}^{2})^{3}\label{eq:tmerge}
\end{equation}
 where $T_{m,0}$ is the merger time in Eq. (\ref{eq:t_merge0-1}),
with $e=0$. The power comes from the fact that the GW decay rate
is $(\dot{a}/a)\propto(1-e^{2})^{-7/2}$, but the binary spends a
fraction of $\sim\sqrt{1-e_{{\rm max}}^{2}}$ of its time near $e\sim e_{{\rm max}}.$
Thus, the actual maximal eccentricity is a crutial parameter in determining
the actual merger times and rates. 

In what follows we derive the analytical result for $e_{{\rm max}}$
in the presence of GR effects and compare to direct simulatioms of
N-body and 2.5PN terms that include gravitational wave inspiral in
the weak field limit.

\subsubsection{Maximal eccentricity}

Taking the total potential to be 
\begin{equation}
\Phi_{{\rm tot}}=\Phi_{{\rm quad}}+\Phi_{{\rm SA}}+\Phi_{{\rm GR}}\label{eq:thitotgr}
\end{equation}
 where the potentials are defined in Eqns. (\ref{eq:phi_quad2}),(\ref{eq:phigr-1}) and (\ref{eq:phicda}).
Similarly to sec. \ref{sec3}, evaluating the total potential at extremal values
of the eccentricity $e_{{\rm max}}$ and $e_{0}\approx0$ leads to
(the full expression for general $e_{0}$ is in Appendix C)

\begin{align}
0 & =\bar{A}\bar{j}_{{\rm min}}^{2}-8\frac{\epsilon_{{\rm GR}}}{\bar{e}_{{\rm max}}^{2}}\bar{j}_{{\rm min}}-15\bar{j}_{z}^{2}\left(1+\frac{9}{8}\epsilon_{{\rm SA}}\bar{j}_{z}\right)\nonumber \\
\bar{A}(\bar{j}_{z},\bar{e}_{{\rm max}}) & \equiv9-\epsilon_{{\rm SA}}\frac{81}{8}\bar{j}_{z}+8\frac{\epsilon_{{\rm GR}}}{\bar{e}_{{\rm max}}^{2}}. \label{eq:jmin_e0}
\end{align}
 For $e_{{\rm max}}\to1,$ Eq. (\ref{eq:jmin_e0}) is a simple quadratic
equation with the solution
\begin{equation}
\bar{j}_{{\rm min}}=\frac{4\epsilon_{{\rm GR}}\pm\sqrt{16\epsilon_{{\rm GR}}^{2}+15\bar{j}_{z}^{2}\left(1+\frac{9}{8}\epsilon_{{\rm SA}}\bar{j}_{z}\right)\bar{A}_{1}}}{\bar{A}_{1}}\label{eq:jmingrapprox}
\end{equation}
where $\bar{A}_{1}=\bar{A}(\bar{j}_{z},\bar{e}_{{\rm max}}\to1, )$.
In the limit of of $\epsilon_{{\rm GR}}=0$ we get back to Eq. (\ref{eq:jzmin}).
In the limit of $\epsilon_{{\rm SA}}=0$ we retain Eq. (52) of \citet[][
since $e_{{\rm max}}\to1$ we implicitly assume $\epsilon_{{\rm GR}}\ll1$]{LML15}. 

In addition, the eccentricity fluctuates by an amount $\delta e$
given be Eq. (\ref{eq:de3}). The actual eccentricity is unbound if
\begin{equation}
\bar{e}_{{\rm max}}^{{\rm SA}}\ge1-\delta e_{{\rm max}}.\label{eq:flip_criteria_with_gr}
\end{equation}
Qualitatively, if $\epsilon_{{\rm GR}}\gg\epsilon_{{\rm SA}}$, the
LK eccentricity oscillations will be quenched and orbital flips will
be suppressed, while for $\epsilon_{{\rm GR}}\ll\epsilon_{{\rm SA}}$,
GR precession is negligible and orbital flips are possible if $\bar{j}_{z}\le\Delta j_{z}$.
In appendix C we show that the critical value that allows unbound
eccentricity and orbital flips is
\begin{equation}
\epsilon_{{\rm GR}}\le\alpha\epsilon_{{\rm SA}}, \label{eq:flipgr}
\end{equation}
where $\alpha=81\sqrt{5/3}/64\approx1.63$. Thus the eccentricity
is unbound if $\bar{j}_{z}\le\Delta j_{z}$ and $\epsilon_{{\rm GR}}\le\alpha\epsilon_{{\rm SA}}$. 

Figure \ref{fig:epsilons} shows the dimensionless parameters $\epsilon_{{\rm GR}}$
and $\epsilon_{{\rm SA}}$ that control the maximal eccentricity.
The left panel shows $\epsilon_{{\rm GR}}$ (solid) and $\epsilon_{{\rm SA}}$
(dashed) on the $a$, $a_{{\rm out}}$ plane, while the right panel
shows the ratio of $\epsilon_{{\rm SA}}/\epsilon_{{\rm GR}}$. Grey
area is the region where $\epsilon_{{\rm GR}}\ge10$, thus eccentricity
excitations are essentially quenched and the binary evolves as an
isolated binary. The purple area is the region of phase space where
the inner binary is unstable to tidal perturbations of the central
object (Hill unstable, $a\ge r_{{\rm H}}=a_{{\rm out}}(m_{{\rm bin}}/3m_{{\rm out}})^{1/3}$).
The pink area in the right panel is the region where $\epsilon_{{\rm GR}}\ge\alpha\epsilon_{{\rm SA}}$;
the maximal eccentricity in bounded, therefore the merger will take
place in the timescale described by Eq. (\ref{eq:tmerge}). Conversely, the white
area allows an unconstrained maximal eccentricity, and thus a direct collision
is possible, given enough time and sufficiently large inclination
(or low $\bar{j}_{z}$).

Fig. \ref{fig:archain} is similar to Fig. \ref{fig:nbody}, but
includes the effects of GR. The initial conditions
are described in the caption. The modified eccentricity is now given
by Eq. (\ref{eq:jmin_e0}), while $\delta e_{{\rm max}}$ is unchanged.
To include effects of GR we use \texttt{ARCHAIN} code \citep{Mikkola06,Mikkola08},
a fully regularized code able to model the evolution of binaries of
arbitrary mass ratios and eccentricities with extreme accuracy, even
over long periods of time. \texttt{ARCHAIN} includes PN corrections up to order
PN2.5, which allows to simulate orbital decay and merger due to GW.

The top panels show realizations for $a_{1}=1,5\ {\rm AU}.$ For $1\ {\rm AU}$,
GR precession is strong enough to quench extreme eccentricity evolution.
Most of the \texttt{ARCHAIN} realizations are slightly below the limiting curve,
with a few orbits with extremely high eccentricities, which are probably
caused by higher order terms in the PN expansion. For $5\ {\rm AU},$
in the region $i_{0}\sim60-80\ {\rm deg}$ (or $i_{0}\sim100-120\ {\rm deg}$
for retrograde cases), additional effects from GR excite the maximal
eccentricity beyond our analytical limit. These effects possibly originate
from higher terms in the PN expansion, or a PN 'interaction term'
(e.g. \citealp{naoz13}) which resonantly enhances the maximal eccentricity.
Additional study of parameter space is presented in appendix D. Studying
these features is beyond the scope of this manuscript and should be
studied elsewhere. In the other regions, where the resonances are
not excited, the maximal eccentricity does follow our analytic prediction.

The bottom panels show realizations for $a_{1}=7,15\ {\rm AU}.$ In
these cases the effects of GR are weak and the maximal eccentricity
behaves similarly to the pure N-body case. In the grey area the eccentricity
is stochastically distributed, such that longer integration times
will decrease $1-e_{{\rm max}}.$ 

\subsubsection{Merger timescale}

Figure \ref{fig:merge_times} shows the maximal eccentricity and merger
times as a function of the inner and outer separations of the systems.
Top to bottom panels show decreasing values of inclination. We see
that increasing inclination decreases the available parameter space
for fast mergers, since the double-averaged maximal eccentricity is
lower. In addition, the overall timescales decrease with increasing
inclination, even though the contours of constant merger time have
a non-trivial structure. 

We can describe the behaviour of equal time curves by the following
heuristic arguments. For a typical example, we look at the curve of
$T_{{\rm m}}=10\ {\rm Gyr}$, corresponding to an initial $a\approx0.2\ {\rm AU}$
at inclination of $i=80^{\circ}$. In the area where $\epsilon_{{\rm GR}}\gg 1 \gg \epsilon_{{\rm SA}}$
the binary is effectively isolated, and the merger time has the same
timescale, regardless of the outer companion. At the point where $\epsilon_{{\rm SA}}\ll\epsilon_{{\rm GR}}\lesssim1$,
the eccentricity is excited, though not to a large value, and the
curve makes a sharp turn to the left. At some point $\epsilon_{{\rm GR}}\sim\epsilon_{{\rm SA}}\ll1$,
so the dependence in $\epsilon_{{\rm GR}}$ weakens, and the changes
in $e_{{\rm max}}$ are less dramatic. The merger time is dominated
by the value of $a$ and the curve takes a turm to the left. At some
point the eccentricity is unbound, so the curve takes a final turn
to the right, and then asymptotically scales with $\epsilon_{{\rm SA}}\propto(a/a_{{\rm out}})^{3/2}$.
The beavior is similar for other contour lines and inclinations. 

To summarize, we have shown that the merger time-scales of hierarchical
triples can be determined analytically given the initial conditions
for systems with comparable mass components where the octupole level of approximation of the triple secular evolution
is suppressed. If the distribution functions of the orbital elements
are known, the fraction and properties of the merging binaries can
be easily estimated and compared to population synthesis simulations;
a subject of future work. 

Note that the limitations on the allowed time for merger can be significantly
shorter than $\sim10\ {\rm Gyr},$ when taking into account dynamical
processes in the Galactic Centre (see \citealp{2012ApJ...757...27A}
and references therein).

\subsection{Formation of Hot Jupiters} \label{42}

\begin{figure*}
\begin{centering}
\includegraphics[width=9cm]{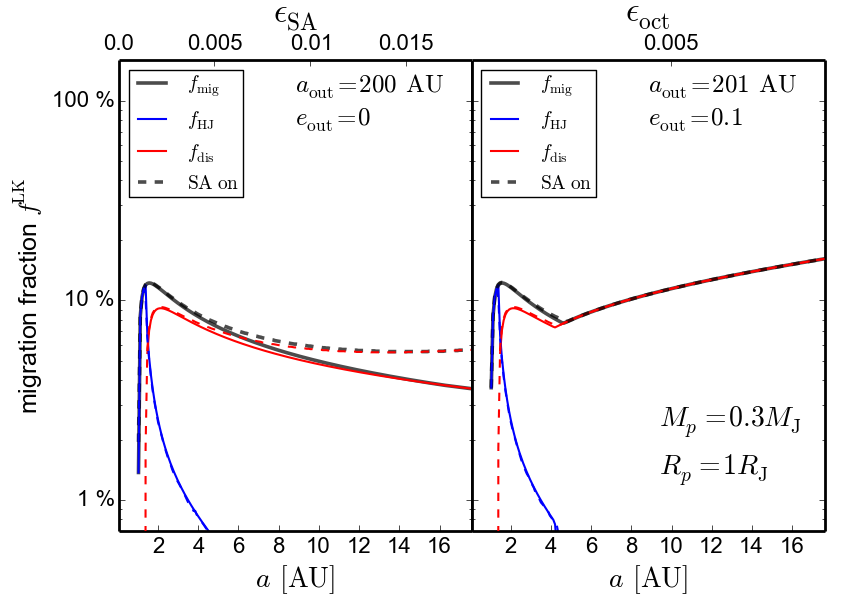}\includegraphics[width=9cm]{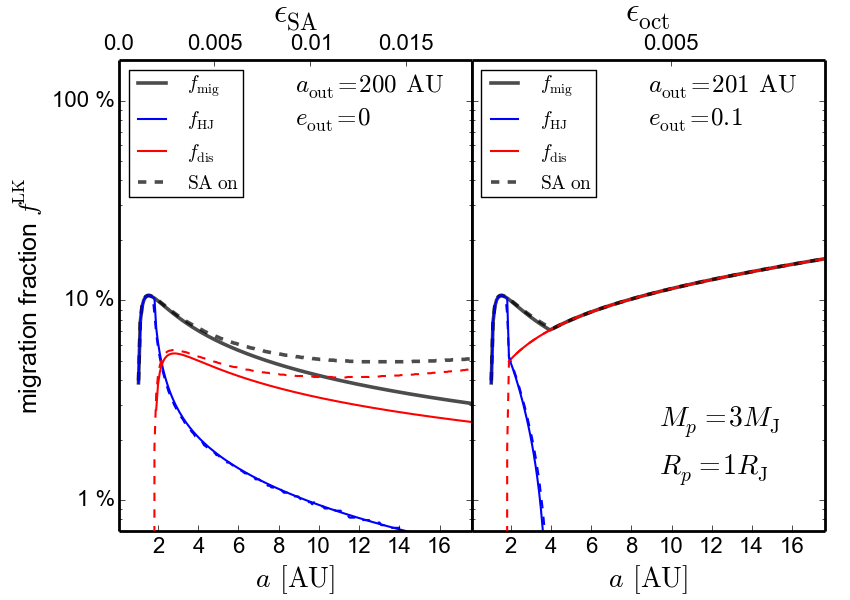}
\par\end{centering}
\caption{\label{fig:3}Hot Jupiter migration, disruption and formation rates
for a system of masses $M_{\star}=1M_{\odot}$, $M_{{\rm out}}=1M_{\odot}$,
with other parameters indicated on the figure. Solid lines are identical
to Figs 2 and 3 from \protect\citet{munoz16}. Dashed lines are the rates with the
inclinaion depends on the corrected maximal eccentricity found in
Eq. (\ref{eq:de3}). Left: Rates for low mass planet of $M_{p}=0.3M_{{\rm J}}.$
Right: Rates for massive planet of $M_{p}=3M_{{\rm J}}.$ Plots reproduced
from publicly available scripts. }
\end{figure*}

Approximate $ \sim 1 \% $ of stars have HJ planets (\citealp{Kn14}; giant planets with period $\lesssim 10 \ \rm{days}$ or semimajor axis $\lesssim 0.1\  \rm{AU}$). It is difficult to form HJs in-situ, therefore dynamical models of LK cycles coupled to tidal friction have been proposed (\citealp{wu03,2007ApJ...669.1298F,2011HJ}; see introduction and sec. \ref{223}). Populations synthesis studies account for only $15-30 \%$ of HJ occurrence rate\footnote{Or higher rates if additional planetary companions are considered \citep{hhj}.} \citep{naoz12, Pet15LKHJ,Anderson_ps}.

Recently, \citet{munoz16} obtained an analytical method for calculating
the migration, disruption and HJ formation rate in terms of the hierarchical
configuration of the planet and the stellar binary. \citet{munoz16} considered a planet of mass $M_{p}$ and radius $R_{p}$ orbiting a
star of mass $M_{\odot}$ with semimajor axis $a$ and eccentrcity $e$, with mutual inclination $i_{\rm tot}$. The star-planet
binary orbits a companion of mass $M_{{\rm \odot}}$, semi-major axis
$a_{{\rm out}}$ and eccentricity $e_{{\rm out}}.$ Similarly to population synthesis models, \citet{munoz16} sampled from uniform distribution the binary properties (e.g. uniform and independent in $\log a_{\rm out}$, $e_{\rm out}$, $a$, $e$,  and $\cos(i_{\rm tot})$) and obtained results which are with population synthesis studies \citep{Pet15LKHJ,Anderson_ps}. The overall fraction of forming HJ is not sensitive to the planetary and stellar physical parameters.

The key parameter is the maximal eccentricity, which is excited by LK oscillations, and suppressed by short-range forces \citep{LML15}. Planets with pericentre
$r_{{\rm disr}}$ (eccentricity $e_{{\rm disr}}$) closer (larger) than
\begin{equation}
r_{{\rm disr}}=a(1-e_{{\rm disr}})=2.7R_{p}\left(\frac{M_{\star}}{M_{p}}\right)^{1/3}\label{eq:rdis}
\end{equation}
 will be disrupted, while planets with pericentre $r_{{\rm mig}}$
(eccentricity $e_{{\rm mig}}$) closer (larger) than 
\begin{align}
r_{{\rm mig}} & =a(1-e_{{\rm mig}})\nonumber \\
 & \approx1.16\left(\frac{Gk_{2p}\tau_{L}}{M_{p}a}\right)^{1/7}M_{\star}^{2/7}R_{p}^{5/7}\tau_{{\rm dis}}^{1/7}(e_{\rm mig}) \label{eq:rmig}
\end{align}
 will migrate within a timescale $\tau_{{\rm dis}}(e_{\rm mig})$, defined in Eq. (\ref{tau_dis_tides}; cf. exact definition in \citealp{munoz16} their Eq. (8) and (9)).  Here $k_{2p}$ is the Love number and $\tau_{L}$ is the lag time discussed in sec. \ref{223}.

Similarly to sec. \ref{22}, \citep{munoz16} found the maximal eccentricity from comparing the total potential 
\begin{equation}
\Phi_{{\rm tot}}=\Phi_{{\rm quad}}+\Phi_{{\rm GR}}+\Phi_{{\rm tide}}, \label{phiall}
\end{equation}
and found the migration, disruption and HJ formation rates for a given binary configuration by taking $f=\arccos(i_{\rm crit})$, where $i_{\rm crit}$ is the critical inclination required to satisfy the disruption or migration radius (Eq. \ref{eq:rdis} and \ref{eq:rmig} respectively), while the HJ formation rate is their difference.

Here we examine how our new formula for the maximal eccentricity changes the results. In our case, we add $\Phi_{\rm SA}$ to the total potential,
 which yields an the implicit equation that determines the maximal eccentricity (in the limit $e_{0}\to0$):
\begin{align}
\frac{3}{5}\bar{j}_{{\rm min}}^{2} & =\bar{j}_{z}^{2}+\frac{8}{15}\frac{\epsilon_{{\rm GR}}}{\bar{e}_{{\rm max}}^{2}}\left(\bar{j}_{{\rm min}}-\bar{j}_{{\rm min}}^{2}\right)\nonumber \\
 & +\frac{8}{225}\frac{\epsilon_{{\rm tide}}}{\bar{e}_{{\rm max}}^{2}}\left(\frac{1+3\bar{e}_{{\rm max}}^{2}+3\bar{e}_{{\rm max}}^{4}/8}{8\bar{j}_{{\rm min}}^{7}}-\bar{j}_{{\rm min}}^{2}\right)\nonumber \\
 & +\epsilon_{{\rm SA}}\frac{9}{8}\bar{j}_{z}\left[\frac{3}{5}\bar{j}_{{\rm min}}^{2}+\bar{j}_{z}^{2}\right]. \label{eq:emaxtide}
\end{align}
The first two terms appear in Eq. (18) of \citet{munoz16}, the last
term is new. 

In order to calculate the migration rates, we use the publicly available
script from \citet{munoz16}\footnote{\href{https://github.com/djmunoz/migration_rates}{https://github.com/djmunoz/migration\_rates}}.
We use the same choice of parameters as in \citet[][ $k_{2p}=0.37,$ $\tau=0.1{\rm \ s},$
$M_{\star}=M_{\odot}$, $M_{p}=M_{J}$, $R_{p}=R_{J}$]{munoz16}, and migration time of $\tau_{{\rm dis}}=1\ {\rm Gyr}$, We change
the prescription for the critical inclination by numerically solving
Eq. (\ref{eq:epstide-1}), for $\bar{e}_{{\rm max}}^{{\rm SA}}\equiv\sqrt{1-\bar{j}_{{\rm min}}^{2}},$
and adding $\delta e_{{\rm max}}$ from Eq. (\ref{eq:de3}).
We then scan the possible grid of inclination range until we find
the critical inclination $i_{c}$ for which $e_{\rm corr} = \bar{e}_{{\rm max}}^{{\rm SA}}(i_{c})+\delta e_{{\rm max}}(i_{c})\ge e_{{\rm crit}}$,
where $e_{{\rm crit}}=e_{{\rm disr}}$ for disruption and $e_{{\rm crit}}=e_{{\rm mig}}$
for migration. There is one to one correspondence between $e_{\rm corr}$ and $i_c$ for prograde inclinations. We tried using retrograde inclinations and got essentially the same results, since the dominating enhancement comes from $\delta e$, which is symmetric in $\cos{i }$.

In Fig. \ref{fig:3} we show the modified rates of the migration,
disruption and HJ formation as a function of the semi-major
axis of the planet. In both panels we see that for closer planets,
the effects of single-averaging are suppressed and the results are
identical to those obtained from the double averaged case. This is
because i) $\epsilon_{{\rm SA}}$ is small and ii) short-range forces
are stronger. As we increase the separation of the planet, the effects
of short-range forces decrease, and $\epsilon_{{\rm SA}}$ increases.
At some point, the SA fraction rates start to diverge from their DA
values and increase. We thus expect an increase of the migration rates
for the outermost planets (or of the least hierarchical system).
Remarkably, both the migration and disruption rates increase, but
the total HJ formation rate is unchanged. This is because
of the narrow range of maximal eccentricities that needs to be satisfied
for efficient HJ formation. 

For eccentric outer binaries, the octupole evolution dominates throughout
the parameter range and the corrected maximal eccentricity has no
effect on the total rates. The caveat is that the detailed octupole
and SA evolution could be different and more work is required in order
to check the consistency of the model. Changing the binary separation
should not significantly affect the results (see Appendix {\bf E}.

To summarize, the total rate of migration and disruption
of HJ increases, but not the total HJ formation rate. Thus, taking
SA effects does not improve the total formation rate of HJ, but it
can increase the rate and observed distributions of migrating warm
Jupiters, with separations of $\sim0.1-1{\rm AU}$ and eccentricity
$e_{p}\ge0.1$. 

\section{Limitations and Caveats}\label{sec5}

\paragraph*{Outer eccentricity and octupole evolution:} Our calculation is exact for circular outer orbits. If the outer eccentricity
is non-zero, i.e. $e_{{\rm out}}\ne0$ then the width of the fluctuations
increases (cf. Eq. \ref{eq:amp}), and the width of the fluctuation
in the eccentricty $\delta e_{{\rm max}}$ is changed (e.g. \citealp{Katz16,HaimKatz18}). The results can be retained for binaries of equal mass.

The more acute effect occurs when the components of the inner binary
have extreme mass ratios. In this case, perturbations from the octupole
term cause a chaotic evolution, and may give rise to extreme eccentricities
and orbital flips \citep{Ford2000,Katz11,EKL-naoz2011, Naoz11nat, li14, Naoz2016review}.
The strength of the octupole term is parametrized by the octupole
parameter $\epsilon_{{\rm oct}}\equiv(a/a_{{\rm out}})e_{{\rm out}}/(1-e_{{\rm out}}^{2})$.

It is unclear how the combined effects of the octupole term and the
single-averaging term change the evolution of the system. On the one
hand, octupole evolution drives the system to a large eccentricity
given sufficiently large mutual inclinations. The larger $\epsilon_{{\rm oct}}$
is, the smaller is the required mutual inclination. Thus, it is expected
that when $\epsilon_{{\rm oct}}\gtrsim\epsilon_{{\rm SA}}$, the octupolar
evolution is restored \citet{Katz16}. In addition, the extra fluctuation
in $j_{z}$ should result in more of the phase space being subjected
to orbital flips and extreme evolution. On the other hand, some systems
do not reach extreme eccentricities, even though the same system could
acquire an orbital flip in the DA approximation (cf. Fig 1. of \citealp{Katz16}).
The actual orbital flip sensitively depends on the initial conditions.
Studying the corrections to the flip criteria in the octupole regime
is beyond the scope of this paper.

\paragraph*{Beyond test particle limit:} We focused on the test particle limit, namely when the outer angular momentum dominates. In the quadrupole approximation, there is still a conserved quantity which depends $j_z$ and the inner-to-outer angular momentum ratio  \citep{katz_sne,naoz13gen,HaimKatz18,LL18}, but the behaviour is more complicated. Recently, \cite{LL18} obtained the maximal eccentricity similarly to sec. \ref{22} and found that the peak inclination for $e_{\rm max} \to 1 $ is shifted to retrograde orbits, and breaks the symmetry. If would be interesting investigate to non-test particle case, a subject of future work.

\paragraph*{N-body integrations:} A major limitation is the finite time of integration. Some of the eccentricities could increase for longer integration times, and the overall stability of the system could be questionable. Most notably, in the area where the orbit could flip, some of the attained eccentricities could be lower. The actual maximal eccentricity in this case could depend on the final time of integration and expected to be distributed similarly to record statistics (N. Haim, private communication). 

\paragraph*{Critical inclination and higher order terms:} Fig. \ref{fig:nbody} shows that for  $\epsilon_{\rm SA}\gtrsim 0.1$ the critical inclination for LK resonance deviates from the predicted value in Eq. (\ref{eq:inc_crit}). The difference is probably from high-order terms in $\epsilon_{\rm SA}$ studied in \citet{cuk04} and \citet{Grishin17}, and/or in additional terms in the multipole expansion that do not vanish for $e_{\rm out}=0$ (cf. Eq. A120 in \citealp{hspz}). In addition, there is a difference of $\sim 15 \ \%$ between the linear estimate of Eq. (\ref{eq:inc_crit}) and the linear estimate in \citet{Grishin17}. We refer to appendix A for discussion and possible solutions.

\section{Summary}\label{sec6}

In this paper we studied the effects of short-term
perturbations on a mildly hierarchical triple system, together with
already studied non-Keplerian perturbations (e.g. general relativity
and tides). We focused on the maximum eccentricity, a key parameter
that determines the result of many short-range interactions and subsequent
evolution of the system. Our result can be summarized as follows:
\begin{enumerate}
\item The strength of the perturbations and typical corrections
are encapsulated in the hierarchy strength (single-averaged, SA) parameter
$\epsilon_{{\rm SA}}$ (Eq. \ref{eq:epssa}).
\item The critical inclinations for the onset of the Lidov-Kozai mechanism change according to Eq. (\ref{eq:inc_crit}), and the overall maximal eccentricity is increased according to Eqns. (\ref{eq:emaxsa}), (\ref{eq:ecorr}) and (\ref{eq:de3}). The new formula is robust, reproduces the orbital flip criteria, is in good agreement with N-body
integrations, and corrects previous work, which has overestimated the eccentricity fluctuations. The main advantage of our calculation is retaining the secular approach, allowing for an efficient computaional approach and better analytic understanding. The double-averaging approximation is not breaking down, but rather is corrected for, such that lesser hierarchies are correctly accounted for. 
\item When general relativistic effects are included, they
tend to add extra precession and quench the secular Lidov-Kozai eccentricity
excitations. We find the maximal eccentriciy with additional general
relativistic precession in Eq. (\ref{eq:jmin_e0}) and the conditions
for orbital flip in Eq. (\ref{eq:flipgr}). We compare to N-body integrations
which include 2.5PN effects and find that our formulae underestimate the maximal eccentricity in some cases, but overall it is in good agreement.
We manifest our results by finding the merger time for black-hole
binaries in the Galactic Centre and argue that the rate or black-hole
mergers due to emission of gravitational waves should be larger when accounting for non-secular effects.
In addition, we found a regime where the maximal eccentricity is unconstrained, direct collisions and/or eccentric mergers of binary black holes are possible, similar to direct collisions of white-dwarfs found by \citet{katz_sne}.
\item We apply our results to hot-Jupiter formation rates.
We include tidal effects and find the maximal eccentricity in this
case in Eq. (\ref{eq:emaxtide}). We incorporate our new maximal eccentricity
in a recent analytical model, and find that the total migration rate
and the disruption rate are increased, but the rate of Hot-Jupiter
formation is unchanged. Nevertheless, the rate for Warm-Jupiter migration
can increase and the underlying observed distributions of migrating
warm Jupiters and their properties are altered, namely Warm-Jupiters
could migrate from further out separations and achieve larger eccentricities.
\end{enumerate}

\section*{Acknowledgements}

We thank Adrian S. Hamers, Dong Lai, Erez Michaely and Diego J. Mu\~noz for discussions and comments on the manuscript. EG acknowledges support from the Technion Irwin and Joan Jacobs Excellence Fellowship for outstanding graduate
students. EG and HBP acknowledge support by Israel Science Foundation I-CORE grant
1829/12 and the Minerva center for life under extreme planetary conditions. GF acknowledges support from a Lady Davis postdoctoral fellowship at the Hebrew University of Jerusalem. GF thanks Seppo Mikkola for helpful discussions on the use of the code \texttt{ARCHAIN}. Simulations were run on the \textit{Astric} cluster at the Hebrew University of Jerusalem.

\bibliographystyle{mnras}
%\bibliography{analytical_e_refs}

\subsection*{Appendix A.\label{appendixA} Critical inclination and Hill stability}

\begin{figure*}
\begin{centering}
\includegraphics[width=7cm]{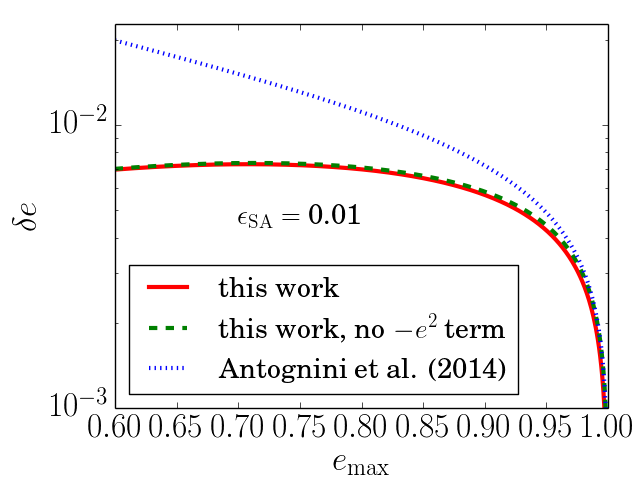}\includegraphics[width=7cm]{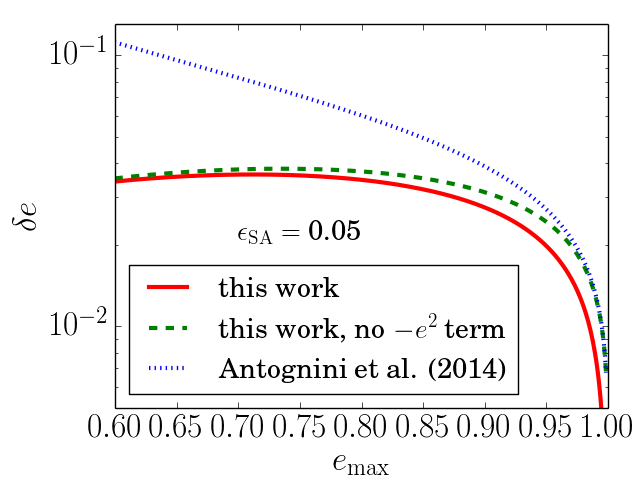}
\par\end{centering}
\begin{centering}
\includegraphics[width=7cm]{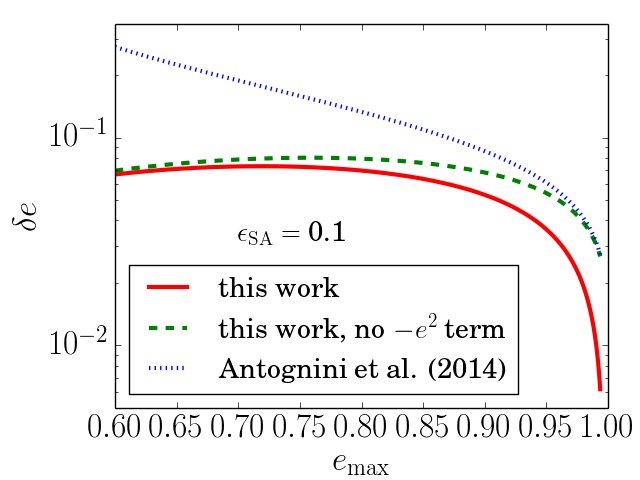}\includegraphics[width=7cm]{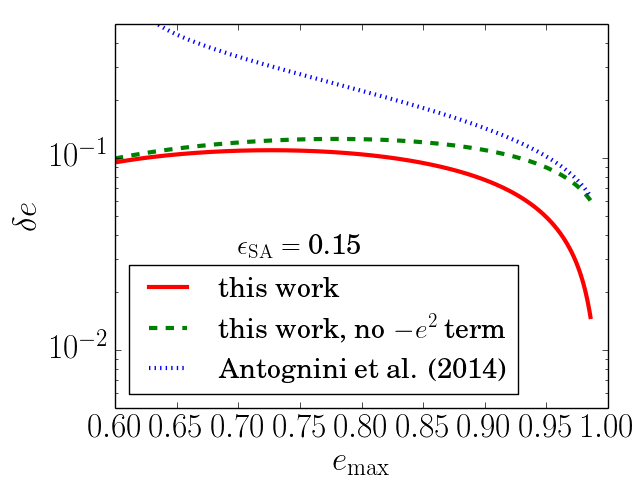}
\par\end{centering}
\caption{\label{fig:antognini}Comparison of our result and \citet{ant14}
for various values of $\epsilon_{{\rm SA}}.$ Each panel shows the
resulting fluctuation as a function of the (double averaged) maximal
eccentricity. Solid red is our Eq. (\ref{eq:de3}), dashed
green is the same equation, but without the last term, dotted blue
is Eq. (\ref{eq:de_antognini1}) (i.e. the corrected version of \protect \citet{ant14}).
In each panel, the maximal eccentricity evaluated is such that $j_{{\rm min}}$
is not smaller than $\Delta j_{z}=9\epsilon_{{\rm SA}}/8,$ otherwise
orbital flip are possible and the eccentricity is unbound. Neglecting
the $\epsilon_{{\rm SA}}e^{2}$ term leads to convergence to \protect \citet{ant14}'s
result for large eccentricity. Generally, \citet{ant14} overestimates
the fluctuation.}
\end{figure*}

For finding the critical inclination, setting $\bar{j}_{{\rm min}}=1^-$ (or $\bar{e}_{\rm max} = 0^+$)
in Eq. (\ref{eq:jzmin}) yields
\begin{align}
1 & =\frac{5}{3}\cos^{2}i_{{\rm crit}}\frac{1+\frac{9}{8}\epsilon_{{\rm SA}}\cos i_{{\rm crit}}}{1-\frac{9}{8}\epsilon_{{\rm SA}}\cos i_{{\rm crit}}}. \label{eq:A1}
\end{align}
This is an implicit equation. Setting $x\equiv\cos i_{{\rm crit}}$
we have an implicit equation
\begin{equation}
F(x,\epsilon_{{\rm SA}})=\frac{5}{3}x^{2}\frac{1+\frac{9}{8}\epsilon_{{\rm SA}}x}{1-\frac{9}{8}\epsilon_{{\rm SA}}x}-1=0, \label{eq:A2}
\end{equation}
 where the local solution at $\epsilon_{{\rm SA}}=0$ is $x_{0}=\sqrt{3/5}$.
Thus, implicit function theorem allows us to get the first order derivative:
\begin{align}
\frac{dx}{d\epsilon_{{\rm SA}}} & =-\left.\frac{\partial F/\partial\epsilon_{{\rm SA}}}{\partial F/\partial x}\right|_{x=x_{0};\ \epsilon_{{\rm SA}}=0} =-\frac{9}{8}x_{0}^{2}=-\frac{27}{40}, \label{eq:51}
\end{align}
thus 
\begin{align}
x(\epsilon_{{\rm SA}}) & =x_{0}+\frac{dx}{d\epsilon_{{\rm SA}}}\epsilon_{{\rm SA}}\nonumber \\
\cos i_{{\rm crit}} & =\sqrt{\frac{3}{5}}-\frac{27}{40}\epsilon_{{\rm SA}}.\label{eq:inc_crita}
\end{align}
 
We can compare with our results in \citet{Grishin17}. Their expansion in the
inclinations is:
\begin{align}
i_{{\rm crit}}=i_{0}+mi_{1}, \label{53}
\end{align}

where $m\equiv\epsilon_{{\rm SA}}$ in the Hill case. The linear correction
is 
\begin{align}
\cos i & =\cos(i_{0}+mi_{1})=\cos i_{0}\cos(mi_{1})-\sin i_{0}\sin(mi_{1}) \nonumber\\
 & =\cos i_{0}-\sin i_{0}i_{1}m, \label{54}
\end{align}
therefore
\begin{align}
i_{1}=\frac{27}{40\sin i_{0}}=\frac{27}{40\sqrt{2/5}}= 1.067\ {\rm rad}, \label{55}
\end{align}
In \citet{Grishin17} we found that the linear correction from Eq.
(10) is $i_{1}=1.24\ {\rm rad},$ and the linear term from the polynomial
fit is $i_{1}=1.17\ {\rm rad}$, which results in errors of $\sim15 \%
$ and $\sim10 \%
,$ respectively. 

The reason is probably lies in the definition of 'LK resonance'. In \citet{Grishin17} we found the inclination for which the pericentre librates, namely the condition $d\omega / dt \approx 0$ on average, while here in deriving Eq. (\ref{eq:inc_crit}) we strictly assumed $\omega = \pi/2$ and looked for a solution for Eq. (\ref{eq:emaxsa}) with $e_{\rm max} = 0^+$. We suspect that Eq. (\ref{eq:inc_crita}) finds the 'bifurcation' where a fixed point in $e-\omega$ space appears near $e\approx 0$, whereas Eq. (10) of \citet{Grishin17} describes where librating solutions are wide spread, and the fixed point is at large eccentricity. Thus, slightly higher inclination is required to satisfy Eq. (10) of \citet{Grishin17}. Future studies may better resolve this issue

\subsection*{B. \label{appendixB}Comparison to Antognini et al. (2014)}

\begin{figure*}
\begin{centering}
\includegraphics[width=8cm]{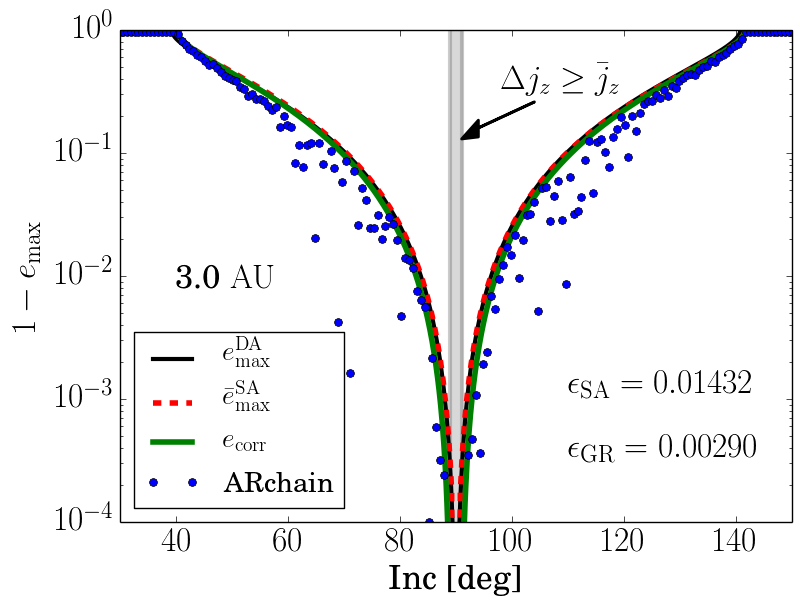}\includegraphics[width=8cm]{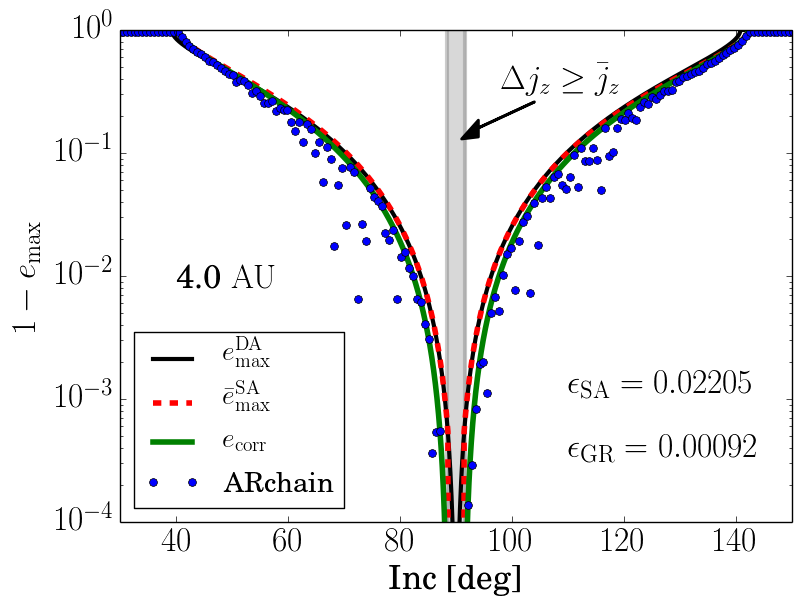}
\par\end{centering}
\begin{centering}
\includegraphics[width=8cm]{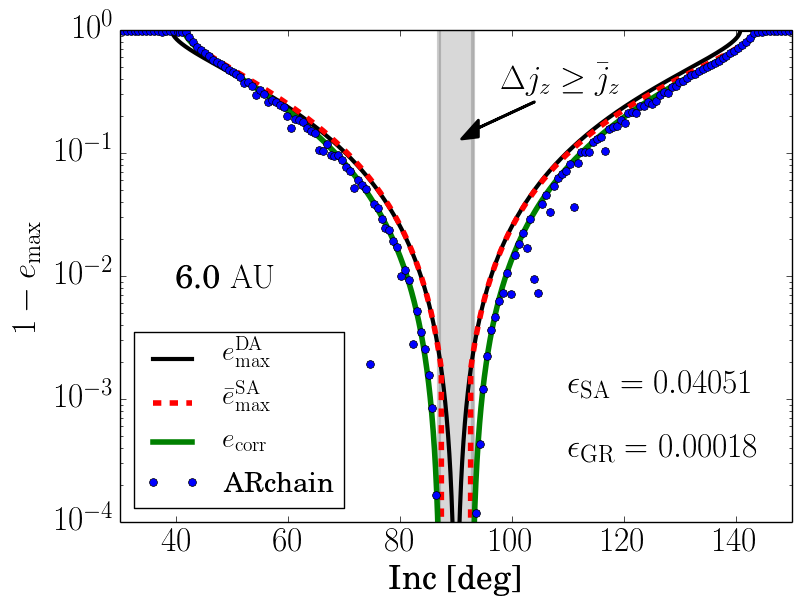}\includegraphics[width=8cm]{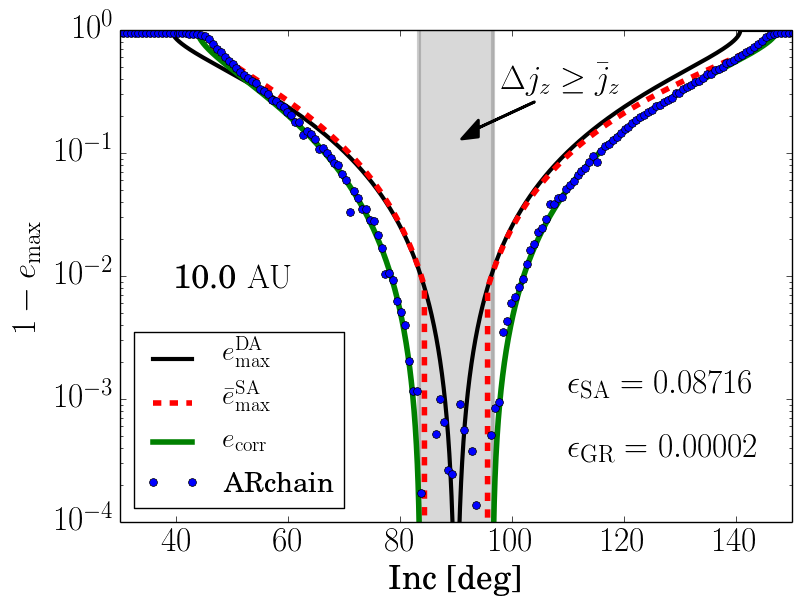}
\par\end{centering}
\caption{\label{fig:archain-1} Same as Fig. \ref{fig:archain}, but with different
inner separations. In each panel, the inner binary separation is $3,4,6$
and $10\ {\rm AU}$ for top left, top right, bottom left and bottom
right, respectively. }
\end{figure*}

Here we compare our formula of the maximal eccentricity
(\ref{eq:de3}) to \citet{ant14}. We start from the result of \citet{Ivanov05}
where the change in the angular momentum near the maximum eccentricity
is
\begin{equation}
\Delta L=\frac{15}{8}\cos i_{{\rm min}}q\left(\frac{a}{a_{{\rm out}}}\right)^{2}\sqrt{Gm_{1}a_{{\rm out}}}, \label{eq:ivaonv1}
\end{equation}
where $q=m_{{\rm out}}/m_{1}$ is assumed to be small. It appears that \citet{ant14}
have confused the mass ratio $q$ in their result, together with additional incorrect prefactors. Note that the change in the normalized
angular momentum $\Delta j$ is 
\begin{equation}
\Delta j=\frac{\Delta L}{L_{{\rm in,0}}}=\frac{15}{8}\cos i_{{\rm min}}\epsilon_{{\rm SA}, }\label{eq:djb}
\end{equation}
 where $L_{{\rm in,0}}=\sqrt{Gm_{1}a}$ and $\epsilon_{{\rm SA}}=q(a/a_{{\rm out}})^{3/2}$
in the limit of $e_{{\rm out}}=0$ and $q\ll1$. Comparing to 
\begin{equation}
\Delta j_{z}=\Delta j\cdot\cos i_{{\rm min}}=\frac{15}{8}\cos^{2}i_{{\rm min}}\epsilon_{{\rm SA}}\label{eq:djzb}
\end{equation}
we get the same result of our Eq. (\ref{eq:de3}) if $\bar{e}_{{\rm max}}\approx1$.
Thus, we expect to converge to \citet{Ivanov05} in this limit. 

The eccentricity is 
\begin{equation}
e=\sqrt{1-\frac{L_{{\rm in}}^{2}}{Gm_{1}a}}, \label{eq:eee}
\end{equation}
which is simply a restatement of the expression for the angular momentum
$L_{{\rm in}}=\sqrt{1-e^{2}}L_{{\rm in,0}}$. The fluctuation is 
\begin{align}
\delta e & =-e+\sqrt{1-\left(\frac{L_{{\rm in}}+\Delta L}{L_{{\rm in,0}}}\right)^{2}}\nonumber \\
 & =-e+\sqrt{1-\left(\sqrt{1-e^{2}}+\frac{15}{8}\cos i_{{\rm min}}\epsilon_{{\rm SA}}\right)^{2}}, \label{eq:deeval}
\end{align}
which is essentially Eq. (3) of \citet{ant14}, up to normalization
factors and corrected mass ratio. Taking $\cos i_{{\rm min}}=\sqrt{3/5}$,
$j_{{\rm min}}\equiv\sqrt{1-e^{2}}$ we get 
\begin{equation}
\delta e=-e+\sqrt{e^{2}-\frac{15}{4}\sqrt{\frac{3}{5}}j_{{\rm min}}\epsilon_{{\rm SA}}-\frac{135}{64}\epsilon_{{\rm SA}}^{2}}.\label{eq:deeval2}
\end{equation}
Note that the correction is in the order of $\mathcal{O}(\epsilon_{{\rm SA}}^{2}),\mathcal{O}(j_{{\rm min}}\epsilon_{{\rm SA}})$,
as expected. Taking $e\approx1$ and expanding the square root we
have 
\begin{align}
\delta e & \approx-e+e\left(1-\frac{15}{8}\sqrt{\frac{3}{5}}j_{{\rm min}}\epsilon_{{\rm SA}}-\frac{135}{128}\epsilon_{{\rm SA}}^{2}\right)\nonumber \\
 & =-\frac{135}{128}\epsilon_{{\rm SA}}\left(\frac{16}{9}\sqrt{\frac{3}{5}}j_{{\rm min}}+\epsilon_{{\rm SA}}\right). \label{eq:de_antognini1}
\end{align}

In the limit of $\bar{e}_{{\rm max}}\approx1$, $j_{{\rm min}}$
is small, thus taking $j_{{\rm min}}=\sqrt{1-\bar{e}_{{\rm max}}}$
leads to a large error, since the fluctuation $\delta j$ could be
comparable to $j_{{\rm min}}$. Indeed, when we compare Eq. (\ref{eq:de_antognini1})
to our Eq. (\ref{eq:de3}), the last term is missing. Taking $j_{{\rm min}}=\bar{j}_{{\rm min}}-\delta j$
where $\delta j$ is given in Eq. (\ref{eq:de_antognini1}) reproduces
the right result in the large $e_{{\rm max}}$/ small $j_{{\rm min}}$
limit. 

Figure \ref{fig:antognini} compares the fluctuations
found in \citet{ant14} (with our corrected version, blue dotted lines),
our work (solid red), and our work without the extra term (dashed
green). Overall, \citet{ant14} overestimate the eccentricity fluctuation.
The error increases with increasing $\epsilon_{{\rm SA}}.$ In \citet{ant14},
their $\epsilon_{{\rm SA}}\approx0.017$, therefore it is hard to
distinguish between different results, although it is evident in their
Fig. 1 that their analytic envelope is indeed overestimating the actual
fluctuations.

\subsection*{C.\label{appendixC} Maximal eccentricity with GR precession}

From the comparison of the total potential (\ref{eq:thitotgr}) we
have for general $e_{0}$:
\begin{align}
 & 9(\bar{e}_{{\rm max}}^{2}-e_{0}^{2})=15\left(\frac{\bar{e}_{{\rm max}}^{2}\bar{j}_{z}^{2}}{\bar{j}_{{\rm min}}^{2}}-e_{0}^{2}\cos^{2}i_{0}\right)+8\epsilon_{{\rm GR}}\left(\frac{1}{\bar{j}_{{\rm min}}}-\frac{1}{j_{0}}\right)\nonumber \\
 & +\frac{27}{8}\epsilon_{{\rm SA}}\bar{j}_{z}\left[3(\bar{e}_{{\rm max}}^{2}-e_{0}^{2})+5\left(\frac{\bar{e}_{{\rm max}}^{2}\bar{j}_{z}^{2}}{\bar{j}_{{\rm min}}^{2}}-e_{0}^{2}\cos^{2}i_{0}\right)\right], \label{eq:phitoteqgr}
\end{align}
 where $j_{0}=\sqrt{1-e_{0}^{2}}$ . 

In the limit of $\bar{j}_{z}\to0,$ $\epsilon_{{\rm GR}}\ll1$ and
$\bar{e}_{{\rm max}}^{{\rm SA}}\to1$ we have 
\begin{align}
\bar{j}_{{\rm min}} & =\frac{4\epsilon_{{\rm GR}}\pm\sqrt{16\epsilon_{{\rm GR}}^{2}+135\bar{j}_{z}^{2}}}{9},\nonumber \\
\delta e_{{\rm max}} & =\frac{135}{128}\epsilon_{{\rm SA}}\left(\frac{16}{9}\sqrt{\frac{3}{5}}|\bar{j}_{{\rm min}}|-\epsilon_{{\rm SA}}\right).\label{eq:jegr-1}
\end{align}
For $\bar{j}_{z}\ll\epsilon_{{\rm GR}}$, we get Eq. (\ref{eq:jegr-1})
for $e_{{\rm lim}}\equiv(1-\bar{j}_{{\rm min}}^{2})^{1/2}\approx1-(32/81)\epsilon_{{\rm GR}}^{2}$
(or $\bar{j}_{{\rm min}}=8\epsilon_{{\rm GR}}/9$). Since the condition
for a prograde-retrograde flip (Eq. \ref{eq:flipgr-1}) is $\bar{j}_{{\rm min}}\lesssim\Delta j_{z}/\cos i_{{\rm min}}=9\sqrt{5/3}\epsilon_{{\rm SA}}/8$,
we plug it in the expression for $\delta e_{{\rm max}}.$ Putting
everything into Eq. (\ref{eq:flip_criteria_with_gr}) we then have
\begin{equation}
\epsilon_{{\rm GR}}\le\alpha\epsilon_{{\rm SA,}}\label{eq:flipgr-1}
\end{equation}
where $\alpha=81\sqrt{5/3}/64\approx1.63$. Thus the eccentricity
is unbound if $\bar{j}_{z}\le\Delta j_{z}$ and $\epsilon_{{\rm GR}}\le\alpha\epsilon_{{\rm SA}}$. 

Note that this derivation demonstrates the need for second order terms is $\delta e_{{\rm max}}$.
A more direct derivation it to compare $\bar{j}_{{\rm min}}=8\epsilon_{{\rm GR}}/9$
from the limiting eccentricity with the fluctuation $\Delta j_{z}/\cos i_{{\rm min}}=9\sqrt{5/3}\epsilon_{{\rm SA}}/8.$

\subsection*{D. \label{appendixD}ARCHAIN realization with additional parameters}

Fig. \ref{fig:archain-1} shows additional realizations of the parameter
space using \texttt{ARCHAIN} 

\subsection*{E.\label{appendixE} HJ formation with different binary separations}

\begin{figure*}
\begin{centering}
\includegraphics[width=9cm]{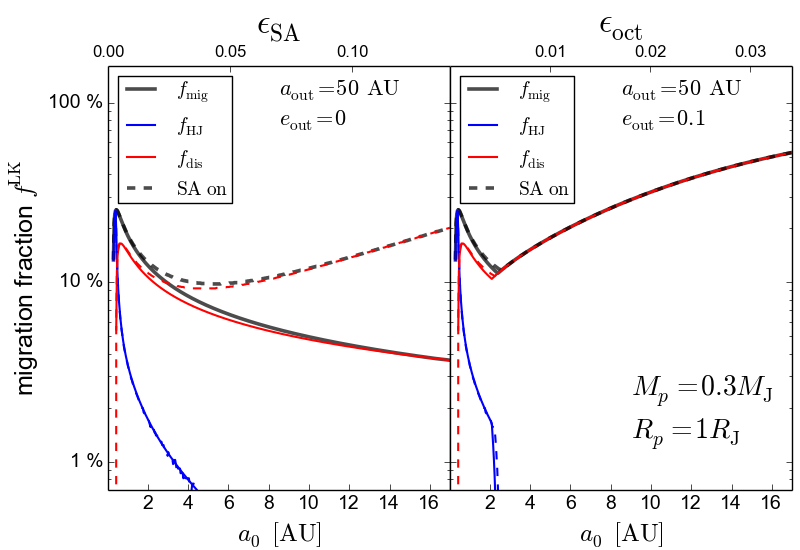}\includegraphics[width=9cm]{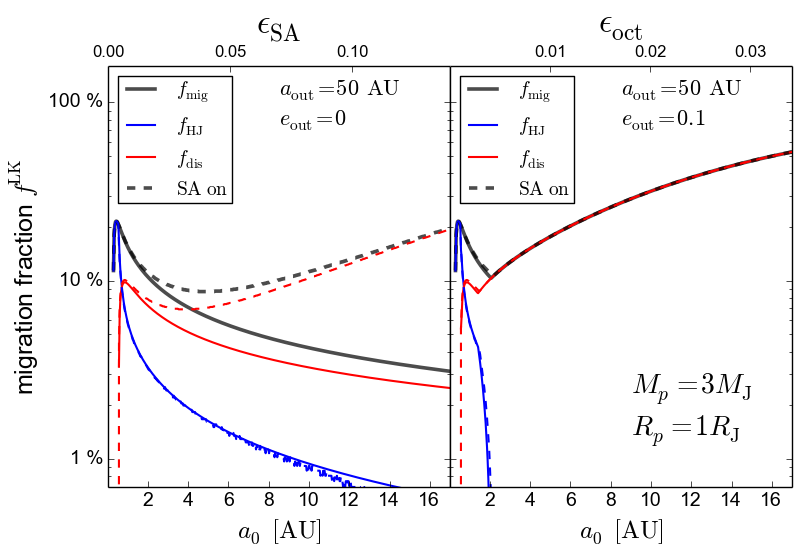}
\par\end{centering}
\caption{\label{fig:3-1} Same as Fig. \ref{fig:3} but with $a_{{\rm out}}=50\ {\rm AU}$. }
\end{figure*}

\begin{figure*}
\begin{centering}
\includegraphics[width=9cm]{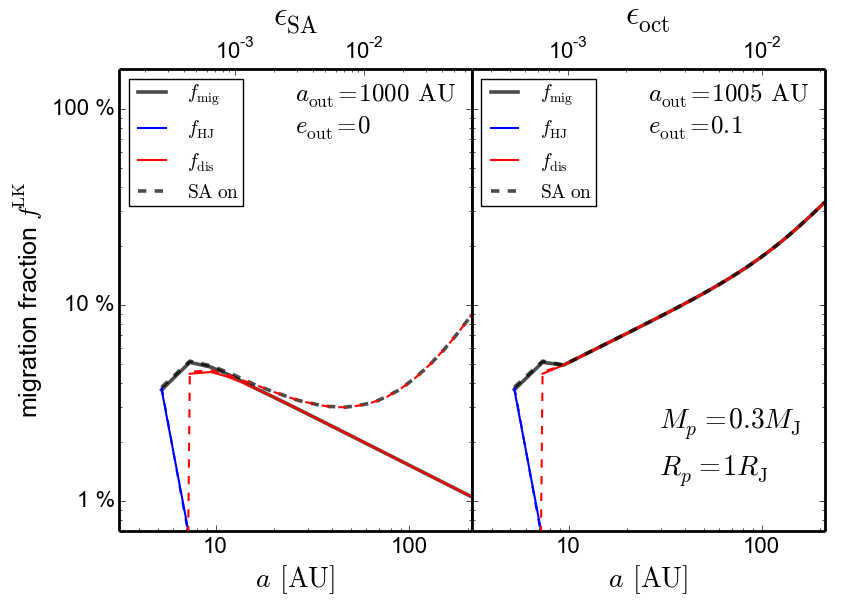}\includegraphics[width=9cm]{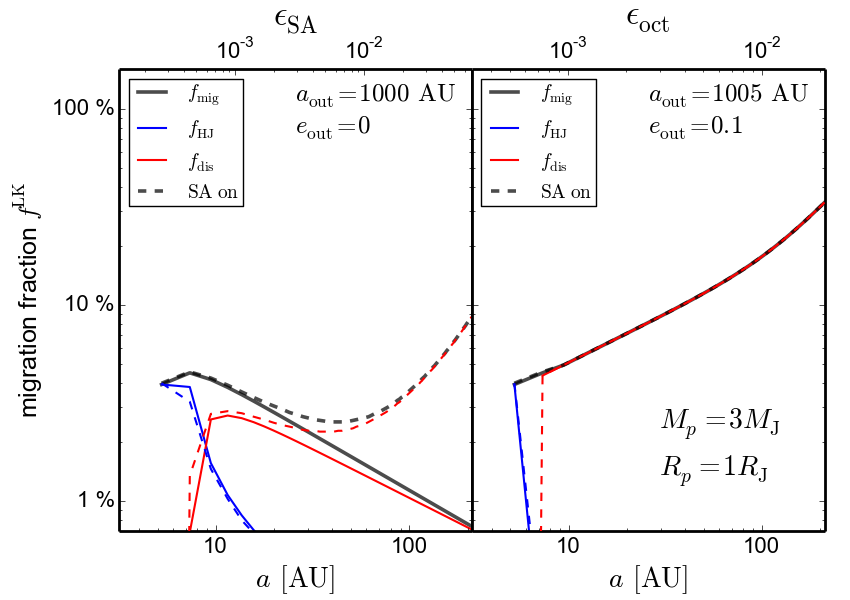}
\par\end{centering}
\caption{\label{fig:3-1-1} Same as Fig. \ref{fig:3} but with $a_{{\rm out}}=1000\ {\rm AU}$.
Note log scale for $x$ axis.}
\end{figure*}
Fig. \ref{fig:3-1} and \ref{fig:3-1-1} show the results of the analytic
model with different values of $a_{{\rm out}}$ and ranges for $a_{{\rm in}}$. 

\end{document}